\documentclass[pdflatex,sn-basic]{sn-jnl}

\usepackage{titlesec}
\usepackage{natbib}
\usepackage{enumitem}
\usepackage{tabularx}
\usepackage{booktabs}
\usepackage{hyperref}
\usepackage{bookmark}
\usepackage{graphicx}%
\usepackage{multirow}%
\usepackage{amsmath,amssymb,amsfonts}%
\usepackage{amsthm}%
\usepackage{bm}
\usepackage{dsfont}
\usepackage{mathrsfs}%
\usepackage[title]{appendix}%
\usepackage{xcolor}%
\usepackage{textcomp}%
\usepackage{manyfoot}%
\usepackage{booktabs}%
\usepackage{algorithm}%
\usepackage{algorithmic}
\usepackage{listings}%
\usepackage{bookmark}
\usepackage{anyfontsize}
\usepackage{tikz}
\usepackage{tikz-3dplot}
\usepackage{colortbl} % in the preamble
\definecolor{lightgray}{gray}{0.9}
\usepackage{makecell}

\renewenvironment{table}[1][]%
{\tableorg[#1]%
\tablebodyfont%
\renewcommand\footnotetext[2][]{{\removelastskip\vskip3pt%
\let\tablebodyfont\tablefootnotefont%
\hskip0pt\if!##1!\else{\smash{$^{##1}$}}\fi##2\par}}%
}{\endtableorg}

\usepackage{float}

\DeclareMathOperator*{\argmin}{arg\,min}

\allowdisplaybreaks

\newtheorem{assumption}{Assumption}

\begin{document}

\title[Article Title]{Gradient Boosting for Spatial Regression Models with Autoregressive Disturbances}

\author*[1]{\fnm{Michael} \sur{Balzer}}\email{michael.balzer@uni-bielefeld.de}

\affil*[1]{\orgdiv{Bielefeld University}, \orgname{Center for Mathematical Economics}, \orgaddress{\street{Universitätsstraße 25}, \city{Bielefeld}, \postcode{33615}, \state{NW}, \country{Germany}}}

\abstract{Researchers in urban and regional studies increasingly deal with spatial data that reflects geographic location and spatial relationships. As a framework for dealing with the unique nature of spatial data, various spatial regression models have been introduced. In this article, a novel model-based gradient boosting algorithm for spatial regression models with autoregressive disturbances is proposed. Due to the modular nature, the approach provides an alternative estimation procedure which is feasible even in high-dimensional settings where established quasi-maximum likelihood or generalized method of moments estimators do not yield unique solutions. The approach additionally enables data-driven variable and model selection in low- as well as high-dimensional settings. Since the bias-variance trade-off is also controlled in the algorithm, implicit regularization is imposed which improves prediction accuracy on out-of-sample spatial data. Detailed simulation studies regarding the performance of estimation, prediction and variable selection in low- and high-dimensional settings confirm proper functionality of the proposed methodology. To illustrative the functionality of the model-based gradient boosting algorithm, a case study is presented where the life expectancy in German districts is modeled incorporating a potential spatial dependence structure.}


\maketitle

\section{Introduction} \label{sec:intro}
In various real-world applications, applied researchers have to engage increasingly with data that exhibits distinct locational attributes, which establishes a connection between objects in space. Such spatial data usually includes a cross-sectional variable in which each observation corresponds to a spatial unit like a geographical location. Thus, dealing with spatial data poses certain challenges since classical statistical regression models heavily rely on the independence assumption between observations which simplifies the model but is commonly violated in the presence of spatial units. Therefore, so-called spatial regression models have been developed which explicitly incorporate the spatial dependence structure of the data \citep{lesage2009}. Particularly, spatial dependencies can arise due to spatial autocorrelation which describes similarities between geographical locations in space. For instance, housing prices in cities might be more similar in neighboring locations of close proximity. Thus, under the assumption that the spatial dependencies only arise through the error terms, \cite{cliff1973} proposed based on \cite{whittle1954} to model the spatial autocorrelation in the disturbance process. Then, for any given variable of interest, the error term for each location depends on a weighted average of the disturbances in connected locations. Thus, spatial spillovers between observations are quantified through the spatial structure of the disturbances \citep{anselin1988}.

Considering that the complexity of the linear predictor can potentially grow quickly in size depending on the modeling choices of the practitioner, approaches to model choice and variable selection have become increasingly more important. Particularly, many candidate spatial regression models with autoregressive disturbances can be chosen by the practitioner by the inclusion of additional spatially lagged independent variables in the linear predictor. However, model choice is non-trivial since modeling spatial dependencies can induce a non-nested model structure. If model estimation is based on the quasi-maximum likelihood (QML), a quantifiable value for model comparison can be obtained based on a likelihood ratio test \citep{lee2004, liu2019}. Instead of utilizing specifically designed criteria, regularization techniques are popular alternatives for model choice and variable selection \citep{fahrmeier2013}. Particularly, the least absolute shrinkage and selection operator \citep{tibshirani1996} is a popular example of such a regularization technique which has also been recently extended to spatial regression model with autoregressive disturbances \citep{cai2019,cai2020}. Another option in classical linear regression models is the so-called model-based gradient boosting algorithm \citep{fahrmeier2013, hepp2016}. Although originally proposed in the domain of machine learning for classification problems, gradient boosting has been extended for statistical regression models where it is known as component-wise, model-based or statistical gradient boosting. In principle, the algorithm is iterative in nature where the estimation problem reduces to fitting base-learners to the negative gradient of a prespecified loss function related to the statistical model of interest. In each iteration of the algorithm only the best performing base-learner is chosen from which a small fraction is added to the current linear predictor. Stopping the algorithm early allows for data-driven variable selection and yields interpretable results at each iteration \citep{mayr2014}. Due to the modular and iterative nature of model-based gradient boosting, it remains a feasible approach even in high-dimensional data settings where the number of variables exceeds the number of observations \citep{bühlmann2006, bühlmann2007}. Therefore, it is not surprising that model-based gradient boosting has been extended to a variety of regression models of increased complexity (see, for example, \cite{jobst2024, balzer2025}).

Although sparse boosting for semi-parametric additive spatial autoregressive models has been recently explored \citep{yue2025}, no prior work appears to have addressed a potential extension of model-based gradient boosting for (parametric) spatial regression models with autoregressive disturbances. To this end, the model-based gradient boosting algorithm in the \textbf{mboost} package \citep{bühlmann2007, hothorn2010, hofner2014, hofner2015} for fitting generalized linear, additive and interaction models of potential high-dimensionality in the programming language R \citep{R} is extended to accommodate spatial regression models with autoregressive disturbances. To investigate proper functionality of the proposed model-based gradient boosting algorithm, in-depth simulation studies in low- and high-dimensional linear settings are conducted. The focus lies primarily on the evaluation of estimation, variable selection and prediction. To illustrate the potential application of model-based gradient boosting for spatial regression models with autoregessive disturbances, an application concerned with modeling the life expectancy in German districts is presented which heavily draws on the "Indikatoren und Karten zur Raum- und Stadtentwicklung" (INKAR) data base which includes a rich variety of variables \citep{BBSR2024}.

The structure in this article is as follows: In Section \ref{sec:meth}, the mathematical, theoretical framework of spatial regression models with autoregressive disturbances and model-based gradient boosting is introduced where both concepts are combined thereafter. Afterward, outcomes for the simulation studies are discussed in Section \ref{sec:sim}. A description of the context, data situation and variables for the case study as well as the results for the application of model-based gradient boosting are presented in Section \ref{sec:case}. The article finishes then with a conclusion and a discussion in Section \ref{sec:con}.

\section{Methodology} \label{sec:meth}
\subsection{Spatial regression models with autoregressive disturbances}
Let $n \in \mathbb{N}$ denote the number of observations in a spatial dataset. For each $i \in \{1,\dots,n\}$, consider following spatial regression model with autoregressive disturbances
\begin{equation} \label{eq:sdem}
\begin{aligned}
    \bm{y} &= \bm{X}\bm{\beta} + \bm{W}\bm{X}\bm{\theta} + \bm{u} \\
    \bm{u} &= \lambda\bm{W}\bm{u} + \bm{\epsilon}
\end{aligned}
\end{equation}
where $\bm{y}$ is a $n \times 1$ vector of observations, $\bm{X}$ is the $n \times p$ design matrix of $p \in \mathbb{N}$ exogenous variables, $\bm{\beta}$ are corresponding $p \times 1$ coefficients and $\bm{u}$ is the $n \times 1$ vector of disturbances. The spatial dependence in the data is assumed to enter the model in two ways. First, the disturbances are modeled as an autoregressive process which depend on a spatial autoregressive parameter $\lambda \in \{-1,1\}$, a spatial weight matrix $\bm{W}$ of size $n \times n$ that captures spatial connections between observations and $n \times 1$ idiosyncratic random innovations $\bm{\epsilon}$. Second, spatial lags of exogenous variables $\bm{W}\bm{X}$ and corresponding $p \times 1$ coefficients $\bm{\theta}$ are also included in modeling the spatial dependent variable of interest. Thus, the model in Equation \ref{eq:sdem} is a generalization of special cases of spatial regression models and is labeled the spatial Durbin error model (SDEM). In the SDEM, spatial autocorrelation is accounted for in both the explanatory variables, as well as the error term. However, removing the spatially lagged independent variables results in the simpler spatial error model (SEM). In contrast, retaining the spatially lagged independent variables but removing the autoregressive nature of the disturbances results in the simpler spatial cross-regressive model (SLX). These special cases are discussed in detail in Appendices \ref{app:sem} and \ref{app:slx} \citep{anselin1988, lesage2009, halleck2015}. 

Additionally, regularity conditions have to be imposed which are useful in ensuring a proper estimation procedure. Assumption \ref{ass:hom} imposes homoskedasticity in the innovations. However, normality is formally not required. In Assumption \ref{ass:W}, the number of observations are linked to the spatial weight matrix which is satisfied if the $\bm{W}$ is row-normalized that is assumed to hold throughout this article. The stability condition $|\lambda| < 1$ ensures invertibility of $\bm{I} - \lambda \bm{W}$ and thus uniqueness of the autoregressive disturbances in terms of innovations. Similarly, Assumption \ref{ass:bound} ensures that the degree of spatial autocorrelation remains manageable \citep{lee2004}. 
\begin{assumption} \label{ass:hom}
    The innovations $\bm{\epsilon}$ are independently and identically distributed with expectation $\mathbb{E}(\epsilon_i) = 0$ and variance $\text{Var}(\epsilon_i) = \sigma^2$. Additionally, for any $\xi > 0$ following moment exists $\mathbb{E}(|\epsilon_1|^{4+\xi}) < \infty$. 
\end{assumption}
\begin{assumption} \label{ass:W}
    The spatial weight matrix $\bm{W}$ contains no self-loops, that is, diagonal entries $w_{ii} = 0$ are set to zero. Off-diagonal entries are $w_{ij} = O\left(\frac{1}{h}\right)$ where $\frac{h}{n} \to 0$.
\end{assumption}
\begin{assumption} \label{ass:non}
    For any $|\lambda| < 1$, the matrix $\bm{I} - \lambda \bm{W}$ exists and is non-singular.
\end{assumption}
\begin{assumption} \label{ass:bound}
    The row and columns sums of $\bm{W}$ and $\left(\bm{I} - \lambda \bm{W}\right)^{-1}$ are uniformly bounded in absolute value.
\end{assumption}

The model in Equation \ref{eq:sdem} can be written more compactly by combining all exogenous variables and the corresponding spatial lags into one design matrix $\bm{Z} =[\bm{X}, \bm{W} \bm{X}]$ and by stacking the corresponding coefficient vertically $\bm{\delta} = (\bm{\bm{\beta}, \bm{\theta}})^{\prime}$ as 
\begin{equation*}
\begin{aligned}
    \bm{y} &= \bm{\eta} + \bm{u} \\
    \bm{u} &=  (\bm{I} - \lambda \bm{W})^{-1} \bm{\epsilon}
\end{aligned}
\end{equation*}
where $\bm{\eta} = \bm{Z} \bm{\delta}$ is the so-called linear predictor. Although not the focus in this article, it is worth to note that the general model formulation also allows for the inclusion of additional spatial lags of exogenous variables, for example, $\bm{W}^2\bm{X}$ or even $\bm{W}^3\bm{X}$. To estimate the coefficients of the exogenous variables and the corresponding spatial lags of exogenous variables, it is convenient to transform the model into a single equation form as
\begin{equation} \label{eq:trans}
    (\bm{I} - \lambda \bm{W})  \bm{y} = (\bm{I} - \lambda \bm{W}) \bm{\eta} + \bm{\epsilon}.
\end{equation}
Then Assumption \ref{ass:col} has to be additionally imposed for a proper estimation procedure which states the usual full column rank condition of $\bm{Z}$.
\begin{assumption} \label{ass:col}
    The limit $\lim_{n \to \infty} \bm{Z}^{\prime}\bm{Z}$ exists, is non-singular and entries of $\bm{Z}$ are uniformly bounded constants.
\end{assumption}

If the autoregressive parameter $\lambda$ is known, then the loss function of Equation \ref{eq:trans} is the squared Mahalanobis distance of the residual vector
\begin{equation}
    \rho(\bm{y}, \bm{\eta}, \bm{\Omega}(\lambda))  = (\bm{y} - \bm{\eta})^{\prime} \bm{\Omega}(\lambda)^{-1} (\bm{y} - \bm{\eta})
\end{equation}
where the error covariance structure is induced by spatial dependence, and
\begin{equation} \label{eq:loss}
    \bm{\Omega}(\lambda) = \sigma^2 \left[(\bm{I} - \lambda \bm{W})^{\prime} (\bm{I} - \lambda \bm{W})\right]^{-1}.
\end{equation}
The negative gradient vector is obtained as the derivative of the loss function with respect to the linear predictor as 
\begin{equation} \label{eq:grad}
    -\frac{\partial}{\partial \bm{\eta}}\rho(\bm{y}, \bm{\eta}, \bm{\Omega}(\lambda))  =  2\bm{\Omega}(\lambda)^{-1} (\bm{y} - \bm{\eta})
\end{equation}
which yields all necessary ingredients for the model-based gradient boosting algorithm \citep{kelejian1999, cai2019, cai2020}.

\subsection{Three-step feasible model-based gradient boosting}
Given the relevant expressions for the ingredients, model-based gradient boosting for the SDEM can be implemented. In principle, the ingredients play an important role in the iterations of the algorithm. Generally, the standard interpretation of boosting as the steepest descent in function space means that the algorithm reduces the empirical risk via the base-learners in an iterative fashion \citep{friedman2001}. The base-learners refer to the functional form of the exogenous input variables. The algorithm begins with an empty model, and fits the specified base-learners to the negative gradient of the chosen loss function. Thus, proper functionality of the algorithm requires a pre-specified loss function which can be quite general. Afterward, the residual sum of squares is computed for each base-learner separately and the linear predictor is updated by a small fraction of the best performing base-learner. The algorithm then reevaluates the negative gradient and updates the linear predictor in an iterative manner until the specified number of boosting iterations are reached \citep{friedman2001, bühlmann2007, mayr2014}. Algorithm \ref{algo:boost} adapts model-based gradient boosting for the SDEM and does not impose any limitations on the number of potential independent variables $q$. Indeed, the great advantage of model-based gradient boosting is the feasibility in high-dimensional settings where the number of variables is larger than the number of available observations. If the squared error is utilized as the loss function, model-based gradient boosting yields a consistent estimator in both low- and high-dimensional settings \citep{zhang2005, bühlmann2006}.

\begin{algorithm}[!htpb] 
\small
\caption{Model-based gradient boosting for the spatial Durbin error model (SDEM)} \label{algo:boost} 
\begin{algorithmic} 
\STATE \textbf{Initialize}
\STATE 1. \textbf{Set} $m = 0$.
\STATE 2. \textbf{Choose} offset values for the linear predictor as $\hat{\bm{\eta}}^{[0]} = (\bm{0})_{\{i = 1, \dots, n\}}$.
\STATE 3. \textbf{Define} the set of base-learners as linear models $\delta_1 \bm{Z}_1, \dots, \delta_q\bm{Z}_q$ where $q$ is the cardinality \\ 
\hspace{0.3cm} of the set of base-learners.
\STATE \textbf{Boosting spatial Durbin error models}
\STATE For $m = 1$ to $m_{\text{stop}}$:
\STATE 4. \textbf{Calculate} the negative gradient vector of the squared Mahalanobis distance at the \\ 
\hspace{0.3cm} current estimates of the linear predictor $\hat{\bm{\eta}}^{[m - 1]}$
    \begin{equation*}
        \bm{v}^{[m]} = \left(v_{i}^{[m]}\right)_{i = 1, \dots, n} = \left( -\frac{\partial}{\partial \bm{\eta}}\rho(\bm{y}, \bm{\eta}, \bm{\Omega}(\lambda)) \Bigg|_{\bm{\eta} = \hat{\bm{\eta}}^{[m - 1]}} \right).
    \end{equation*}
\STATE 5. \textbf{Fit} each specified linear model base-learner to the negative gradient vector separately.
\STATE 6. \textbf{Evaluate} the residual sum of squares for each base-learner and pick component $j^{*}$ that \\ 
\hspace{0.3cm} fits the negative gradient vector best
    \begin{equation*}
        j^{*} = \operatorname*{argmin}_{1 \leq j \leq q} \sum_{i=1}^{n} \left(v^{[m]}_i - \hat{\delta}^{m}_{j}z_j\right)^2.
    \end{equation*}
\STATE 7. \textbf{Update} the linear predictor $\hat{\bm{\eta}}$ based on $j^{*}$ according to 
\begin{equation*}
        \hat{\bm{\eta}}^{[m]} = \hat{\bm{\eta}}^{[m - 1]} + s \cdot \hat{\delta}^{m}_{j^{*}} z_{j^{*}}
    \end{equation*}
\\ \hspace{0.3cm} where $s$ is a small learning rate or step-length.
\end{algorithmic}
\end{algorithm}

Additionally, \cite{bühlmann2007} formally show that model-based gradient boosting with squared error loss converges to the ordinary least squares (OLS) solution if the number of boosting iterations $m_{\text{stop}}$ is chosen sufficiently large. As the main tuning parameter in the algorithm, $m_{\text{stop}}$ controls the so-called bias-variance trade-off. Thus, the accuracy of prediction can be improved and overfitting behavior mitigated. Due to the modular nature, the algorithm yields an interpretable solution at each iteration such that sparser models can be obtained by stopping the algorithm early instead of convergence. Since the algorithm only updates the linear predictor by one component at each iteration, variable selection and effects estimation shrinkage is also accounted for \citep{mayr2012}.

The feasibility of Algorithm \ref{algo:boost} strongly relies on the assumption that the spatial autoregressive parameter $\lambda$ is apriori known. However, in real-world application settings, $\lambda$ is unknown which implies that $\bm{\Omega}(\lambda)$ occurring in the squared Mahalanobis distance and the negative gradient cannot be evaluated. Therefore, a three-step model-based gradient boosting procedure is proposed to enable the feasibility of Algorithm \ref{algo:boost} which relies on replacing the unknown $\lambda$ and $\bm{\Omega}(\lambda)$ by the estimated counterparts $\hat{\lambda}$ and $\bm{\Omega}(\hat{\lambda})$. In the first step, the model in Equation \ref{eq:sdem} is written as 
\begin{equation} \label{eq:sim}
    \bm{y} = \bm{Z}\bm{\delta}+ \bm{u}
\end{equation}
temporarily ignoring the potential autoregressive structure of the disturbances. The model in Equation \ref{eq:sim} can then be estimated using a variety of methods, as long as the resulting estimator $\bm{\tilde{\delta}}$ is consistent. For low-dimensional settings, a natural choice for the estimator is OLS. However, since the loss function associated with Equation \ref{eq:sim} is simply the squared error loss, model-based gradient boosting can also be employed. In high-dimensional settings, OLS does not yield unique solutions, so model-based gradient boosting which provides a consistent estimator is utilized instead.

In the second step, let $\bm{\tilde{u}} = \bm{y} -\bm{Z}\bm{\tilde{\delta}}$ denote the predictors of $\bm{u}$ based on a consistent estimator $\bm{\tilde{\delta}}$. Define $\bm{\bar{u}} = \bm{W}\bm{u}$, $\bm{\bar{\bar{u}}} = \bm{W}\bm{W}\bm{u}$ and the corresponding expressions based on the predictors as $\bm{\tilde{\bar{u}}} = \bm{W}\bm{\tilde{u}}$ and $\bm{\tilde{\bar{\bar{u}}}} = \bm{W}\bm{W}\bm{\tilde{u}}$. Adapt an identical notation pattern to $\bm{\bar{\epsilon}} = \bm{W} \bm{\epsilon}$. Then, if Assumptions \ref{ass:hom} to \ref{ass:non} hold, following three moments can be obtained
\begin{equation} \label{eq:mom}
    \mathbb{E}\left(\frac{1}{n} \bm{\epsilon}^{\prime}\bm{\epsilon} \right) = \sigma^2 \quad \mathbb{E}\left(\frac{1}{n} \bm{\bar{\epsilon}}^{\prime}\bm{\bar{\epsilon}} \right) = \sigma^2 \frac{1}{n} \text{tr}\left(\bm{W}^{\prime}\bm{W}\right) \quad   \mathbb{E}\left(\frac{1}{n} \bm{\bar{\epsilon}}^{\prime}\bm{\epsilon} \right) = 0
\end{equation}
where $\text{tr}(\cdot)$ denotes the trace of any matrix. Since the innovations can be written in terms of $\bm{\bar{u}}$ and $\bm{\bar{\bar{u}}}$ as $\bm{\epsilon} = \bm{u} - \lambda\bm{\bar{u}}$ and $\bm{\bar{\epsilon}} = \bm{\bar{u}} - \lambda\bm{\bar{\bar{u}}}$, a system of three equations can be obtained based on Equations \ref{eq:sdem} and \ref{eq:mom}
\begin{equation} \label{eq:sol}
    \bm{\Gamma}[\lambda, \lambda^2, \sigma^2]^{\prime} - \bm{\gamma} = 0.
\end{equation}
The expressions for $\bm{\Gamma}(\cdot)$ and $\bm{\gamma}$ are given as 
\begin{equation*}
    \bm{\Gamma} =
\begin{bmatrix}
\frac{2}{N} \bm{u}^\top \bm{\bar{u}} & -\frac{1}{N} \bm{\bar{u}}^\top \bm{\bar{u}} & 1 \\
\frac{2}{N} \bm{\bar{\bar{u}}}^\top \bm{\bar{u}} & -\frac{1}{N} \bm{\bar{\bar{u}}}^\top \bm{\bar{\bar{u}}} & \frac{1}{N} \operatorname{tr}(\bm{W}^\top \bm{W}) \\
\frac{1}{N} \left( \bm{u}^\top \bm{\bar{\bar{u}}} + \bm{\bar{u}}^\top \bm{\bar{u}} \right) & -\frac{1}{N} \bm{\bar{u}}^\top \bm{\bar{\bar{u}}} & 0
\end{bmatrix} \quad \bm{\gamma} =
\begin{bmatrix}
\frac{1}{N} \bm{u}^\top \bm{u} \\
\frac{1}{N} \bm{\bar{u}}^\top \bm{\bar{u}} \\
\frac{1}{N} \bm{u}^\top \bm{\bar{u}}
\end{bmatrix}.
\end{equation*}
Replacing the moments in Equation \ref{eq:sol} by the corresponding sample moments yields
\begin{equation} \label{eq:est}
    \bm{G}[\lambda, \lambda^2, \sigma^2]^{\prime} - \bm{g} = \bm{\nu}(\lambda, \sigma^2).
\end{equation}
where 
\begin{equation*}
    \bm{G} =
\begin{bmatrix}
\frac{2}{N} \bm{\tilde{u}}^\top \bm{\tilde{\bar{u}}} & -\frac{1}{N} \bm{\tilde{\bar{u}}}^\top \bm{\tilde{\bar{u}}} & 1 \\
\frac{2}{N} \bm{\tilde{\bar{\bar{u}}}}^\top \bm{\tilde{\bar{u}}} & -\frac{1}{N} \bm{\tilde{\bar{\bar{u}}}}^\top \bm{\tilde{\bar{\bar{u}}}} & \frac{1}{N} \operatorname{tr}(\bm{W}^\top \bm{W}) \\
\frac{1}{N} \left( \bm{\tilde{u}}^\top \bm{\tilde{\bar{\bar{u}}}} + \bm{\tilde{\bar{u}}}^\top \bm{\tilde{\bar{u}}} \right) & -\frac{1}{N} \bm{\tilde{\bar{u}}}^\top \bm{\tilde{\bar{\bar{u}}}} & 0
\end{bmatrix} \quad \bm{g} =
\begin{bmatrix}
\frac{1}{N} \bm{\tilde{u}}^\top \bm{\tilde{u}} \\
\frac{1}{N} \bm{\tilde{\bar{u}}}^\top \bm{\tilde{\bar{u}}} \\
\frac{1}{N} \bm{\tilde{u}}^\top \bm{\tilde{\bar{u}}}
\end{bmatrix}
\end{equation*}
and $\bm{\nu}(\lambda, \sigma^2)$ is interpreted as a $3 \times 1$ residual vector. The estimators for $\lambda$ and $\sigma^2$ are obtained using nonlinear least squares and are denoted by $\hat{\lambda}$ and $\hat{\sigma}^2$. Based on Equation \ref{eq:est}, the non-linear least squares estimators are defined as
\begin{equation}
(\hat{\lambda}, \hat{\sigma}^2) = \argmin_{\lambda, \sigma^2} \left[\bm{G}[\lambda, \lambda^2, \sigma^2]^{\prime} - \bm{g} \right]^{\prime} \left[\bm{G}[\lambda, \lambda^2, \sigma^2]^{\prime} - \bm{g} \right].
\end{equation}
Let the Assumptions \ref{ass:hom} to \ref{ass:away} hold. Then the nonlinear least squares estimators $\hat{\lambda}$ and $\hat{\sigma}^2$ are consistent estimators of $\lambda$ and $\sigma^2$ in the sense that $\hat{\lambda} \to_p \lambda$ and $\hat{\sigma}^2 \to_p \sigma^2$ for $n \to \infty$ sufficiently large.
\begin{assumption} \label{ass:fin}
Denote by $\tilde{u}_{i}$ denote the $i$-th element of the vector $\bm{\tilde{u}}$.  
Assume that there exist finite-dimensional random vectors $\bm{d}_{i,n}$ and $\bm{\Delta}_n$ such that
\[
\left| \tilde{u}_{i} - u_{i} \right| \leq \left\| \bm{d}_{i,n} \right\| \left\| \bm{\Delta}_n \right\|
\]
with the following conditions holding
\[
\frac{1}{n} \sum_{i=1}^n \left\| \bm{d}_{i,n} \right\|^{2 + \delta} = O_p(1) \quad \text{for some } \delta > 0, \quad \text{and} \quad \sqrt{n} \left\| \bm{\Delta}_n \right\| = O_p(1).
\]
\end{assumption}
\begin{assumption} \label{ass:away}
Assume that the matrix $\bm{\Gamma}^{\prime} \bm{\Gamma}$ is well-conditioned in the sense that its smallest eigenvalue is bounded away from zero. Specifically,
\[
\phi_{\min} \left( \bm{\Gamma}_n^{\prime} \bm{\Gamma}_n \right) \geq \phi^{\ast} > 0,
\]
where the constant $\phi^{\ast}$ may depend on $\lambda$ and $\sigma^2$.
\end{assumption}

In the third step, replace $\lambda$ in $\bm{\Omega}(\lambda)$ by the estimated counterparts $\hat{\lambda}$ yielding $\bm{\Omega}(\hat{\lambda})$. Thus, the squared Mahalanobis distance and negative gradient become
\begin{align} \label{eq:ende}
        \rho(\bm{y}, \bm{\eta}, \bm{\Omega}(\hat{\lambda}))  &= (\bm{y} - \bm{\eta})^{\prime} \bm{\Omega}(\hat{\lambda})^{-1} (\bm{y} - \bm{\eta}) \\
         -\frac{\partial}{\partial \bm{\eta}}\rho(\bm{y}, \bm{\eta}, \bm{\Omega} \label{eq:ende2}(\hat{\lambda}))  &=  2\bm{\Omega}(\hat{\lambda})^{-1} (\bm{y} - \bm{\eta}) \\
             \bm{\Omega}(\hat{\lambda}) &= \hat{\sigma}^2 \left[(\bm{I} - \hat{\lambda}\bm{W})^{\prime} (\bm{I} - \hat{\lambda} \bm{W})\right]^{-1}.
\end{align}
Finally replace the expressions in Algorithm \ref{algo:boost} by the corresponding expressions based on the estimated counterparts in Equations \ref{eq:ende} and \ref{eq:ende2} which yields a feasible model-based gradient boosting algorithm for the SDEM \citep{kelejian1999}.

\subsection{Post-hoc deselection}
In model-based gradient boosting, the standard approach for model and variable selection is through early stopping via the stopping criterion $m_{\text{stop}}$. Generally, $m_{\text{stop}}$ is usually chosen by means of k-fold cross-validation, subsampling or bootstrapping which have the tendency to include to many (non-informative) variables which a induces non-parsimonious final model \citep{mayr2012}. Since the quality of the generalized moment estimators $\hat{\lambda}$ and $\hat{\sigma}^2$ depend on the quality of the predictors of $\bm{u}$, the inclusion of non-informative variables severely impacts the final estimators.

To mitigate the consequences and obtain sparser final models, the deselection algorithm proposed in \cite{stromer2022} is adapted to model-based gradient boosting for the SDEM which can be easily utilized in the first and third step of the procedure. In principle, the idea is to perform model-based gradient boosting and determine the optimal stopping iteration $m_{\text{opt}}$ using standard cross-validation techniques. Subsequently, variables that contribute the least to risk reduction are deselected, and model-based gradient boosting is reapplied using the remaining variables and the previously determined $m_{\text{opt}}$. More formally, let $\mathds{1}(\cdot)$ denote the indicator function, then the attributable risk reduction given by 
\begin{equation}
    R_j = \sum_{m = 1}^{m_{\text{stop}}} \mathds{1}\left\{j = j^{{*}^{[m]}}\right\} \left(r^{[m-1]} - r^{[m]}\right), \quad j = 0,1,2,\dots, q
\end{equation}
is assumed to be a measure of importance for the $j$-th base-learner. Then $j^{{*}^{[m]}}$ is the component corresponding to the selected base-learner, $\left(r^{[m-1]} - r^{[m]}\right)$ is the risk reduction in iteration $m$ where $r^{[m-1]}$ and $r^{[m]}$ are corresponding risks. A base-learner and thereby the corresponding variable is deselected if 
\begin{equation} \label{eq:des}
    R_j < \tau \left(r^{[0]} - r^{[m_{\text{stop}}]}\right)
\end{equation}
where $\tau \in (0,1)$ is a pre-specified threshold and $\left(r^{[0]} - r^{[m_{\text{stop}}]}\right)$ the total risk reduction. Thus, variables remain only in the model if the relative risk contribution is equal or larger than the threshold $\tau$. The choice of $\tau$ usually depends on the particular research situation at hand. However, a typical recommendation is to keep values low, that is, for example, $\tau = 0.01$ \citep{stromer2022}. The complete procedure incorporating the deselection approach into model-based gradient boosting for the SDEM is given in Algorithm \ref{algo:des}.
\begin{algorithm}[!htpb] 
\small
\caption{Model-based gradient boosting for the spatial Durbin error model (SDEM) with additional deselection of variables} \label{algo:des} 
\begin{algorithmic} 
\STATE 1. \textbf{Initialization:} Estimate the SDEM via model-based gradient boosting with an initial \\ \hspace{0.3cm} number of boosting iterations $m_{\text{stop}}$ according to Algorithm \ref{algo:boost}.
\STATE 2. \textbf{Optimization:} Search for the optimal stopping criterion $m_{\text{opt}}$ via k-fold cross- \\ \hspace{0.3cm} validation, subsampling or bootstrapping.
\STATE 3. \textbf{Deselection:} Deselect variables according to the lowest impact on the risk reduction \\ \hspace{0.3cm} given by Equation \ref{eq:des}.
\STATE 4. \textbf{Finalization:} Estimate the SDEM via model-based gradient boosting with the \\ \hspace{0.3cm} remaining variables and optimal stopping criterion $m_{\text{opt}}$ according to Algorithm \ref{algo:boost} again.
\end{algorithmic}
\end{algorithm}

\section{Simulation Study} \label{sec:sim}
\subsection{Study Design}
To evaluate the performance of the proposed three-step model-based gradient boosting algorithm, simulation studies are conducted with the ingredients derived in Section \ref{sec:meth}. Particularly, the performance of estimation, variable selection and prediction in low- as well as high-dimensional linear settings are evaluated. Additionally, an evaluation of the performance of the deselection algorithm is provided. The study design is motivated by the case study of modeling life expectancy in German districts in Section \ref{sec:case}. Specifically, the number of observations is fixed at $n = 400$. In contrast, the number of independent variables 
is varied between $q = 20$ and $q = 800$, indicating a low- $(n > q)$ and high-dimensional $(n < q)$ linear setting. The true data generating process is given by 
\begin{equation*}
\begin{aligned}
    \bm{y} &= 1 + 3.5\bm{X}_1 -2.5 \bm{X}_2 -4 \bm{W} \bm{X}_1 + 3 \bm{W} \bm{X}_2 + \bm{u} \\
    \bm{u} &= \lambda\bm{W}\bm{u} + \bm{\epsilon}
\end{aligned}
\end{equation*}
where the variables are independently and identically drawn from the uniform distribution $\bm{X} \sim U(-2,2)$. The spatial autoregressive parameter is varied throughout the simulation study by $\lambda \in \{-0.8,-0.6,-0.4,-0.2,0.2,0.4,0.6,0.8\}$ and the innovations are normally distributed according to $\bm{\epsilon} \sim N(0,\sigma^2)$ with $\sigma^2 =1$. The spatial weight matrix $\bm{W}$ is generated based on a circular world in which each location is directly related to the five locations before and after, that is, $K = 5$. Additionally, $\bm{W}$ is row-normalized such that each row sums up to one. Simulation studies for varying number of related locations $K \in \{1,2,3,5,10,20\}$ is given in Appendix \ref{app:mat}. In the model-based gradient boosting algorithm, the corresponding base-learners are specified as simple linear regression models due to the nature of the data generating process. The learning rate is set to $s = 0.1$ since that is the usual practice (see, for example, \cite{schmid2008, mayr2012, hofner2014}). The optimal stopping criterion $m_{\text{opt}}$ is found by minimizing the empirical risk via 25-fold subsampling. In each simulation setting, a total of $n_{\text{sim}} = 100$ repetitions are conducted. Additionally, different approaches for the consistent estimator $\bm{\tilde{\delta}}$ in the first step are considered, along with their impact on the final results. Specifically, the reported methods are first-step OLS (LS-GB), first-step gradient boosting (GB-GB), and first-step gradient boosting with deselection (DS-GB) where applicable.

Regarding the performance of variable selection and deselection, the criteria are chosen based on the confusion matrix. In particular, the reported variable selection criteria are the true positive rate (TPR), which is the proportion of correctly selected variables out of all true informative variables, the true negative rate (TNR), which is the proportion of correctly non-selected variables out of all true non-informative variables and the false discovery rate (FDR), which is the proportion of non-informative variables in the set of all selected variables \citep{stehmann1997}.

The performance of estimation is evaluated by reporting the bias, the mean squared error (MSE) and the empirical standard error (ESE) for $\lambda$ defined as
\begin{align*}
    \text{Bias} &= \frac{1}{n_{\text{sim}}} \sum_{i = 1}^{n_{\text{sim}}} \hat{\lambda}_i - \lambda \\
    \text{MSE} &= \frac{1}{n_{\text{sim}}} \sum_{i = 1}^{n_{\text{sim}}} (\hat{\lambda}_i - \lambda)^2 \\
    \text{ESE} &= \sqrt{\frac{1}{n_{\text{sim}} - 1} \sum_{i = 1}^{n_{\text{sim}}} (\hat{\lambda}_i - \bar{\lambda})^2}.
\end{align*}
For all proposed performance criteria, lower values are always preferred. Additionally, the effects shrinkage estimation for the independent variables is evaluated via visualization by boxplots to highlight the median, outliers and quartiles over 100 repetitions \citep{morris2019}.

Furthermore, the prediction accuracy is evaluated based on an additional test data set. The test data $\bm{y_{\text{test}}}$ is generated according to the same data generating process as the train data with an identical number of observations $n_{\text{test}} = 400$. The chosen criteria are the quasi negative log-likelihood (NLL), the root mean squared error of prediction (RMSEP) and mean absolute error of prediction (MAEP) defined as
\begin{align*}
     \text{RMSEP} &= \sqrt{ \frac{1}{n_{\text{test}}} \sum_{i=1}^{n_{\text{test}}} (y_{\text{test},i}  - \hat{y}_{\text{test},i})^2} \\
     \text{MAEP} &=  \frac{1}{n_{\text{test}}} \sum_{i=1}^{n_{\text{test}}} |y_{\text{test},i} - \hat{y}_{\text{test},i}| \\
     \text{NLL} &=  
    \frac{n_{\text{test}}}{2} \left(\log(2\pi\hat{\sigma}) + 1 \right)
    - \log \left| I - \hat{\lambda} W \right| \\
    &+ \frac{1}{2\sigma} (\bm{y_{\text{test}}} - \bm{\hat{\eta}_{\text{test}}})^{\prime} (I - \hat{\lambda} W)^{\prime} (I - \hat{\lambda} W)(\bm{y_{\text{test}}} - \bm{\hat{\eta}_{\text{test}}})
\end{align*}
For all proposed performance criteria of prediction accuracy, lower values are always preferred.

The simulation study is conducted in the programming language \textbf{R} \citep{R}.
The QML and generalized method of moments (GMM) estimation of the SDEM in the low-dimensional linear setting is performed via the \textbf{spatialreg} package \citep{bivand2021, pebesma2023}. The presented graphics are created with the \textbf{tidyverse} packages \citep{tidyvere2019}. Model-based gradient boosting for generalized, additive and interaction models can be found in the \textbf{mboost} package \citep{bühlmann2007, hothorn2010, hofner2014, hofner2015}. An implementation for model-based gradient boosting for the SDEM via the novel spatial error family incorporating the deselection algorithm and the R code for reproducibility of all simulation studies can be found in the GitHub repository \url{https://github.com/micbalz/SpatRegBoost}.

\subsection{Results}
\subsubsection{Low-dimensional linear setting (n = 400, p = 20)}
In Table \ref{tab:sel}, the average selection rates over 100 repetitions for the low-dimensional linear settings can be seen. The results show a consistent TPR of $100\%$ across all spatial autoregressive parameters $\lambda$, indicating that all informative variables are selected on average. In contrast, for negative values of $\lambda$, a TNR of around $25\%$ is achieved, indicating that the method correctly avoids selecting only about four out of 16 non-informative variables on average, and incorrectly selects approximately 12 of them. As $\lambda$ becomes more positive, the TNR steadily increases, reaching its highest value of around $75\%$ for $\lambda = 0.8$, meaning that only about four out of 16 non-informative variables are incorrectly selected on average. However, a low TNR is accompanied by a high FDR across all values of $\lambda$. Although the TNR steadily increases as $\lambda$ becomes more positive, the corresponding decrease in FDR is less pronounced. Even at $\lambda = 0.8$, where the TNR is highest, the FDR remains substantial at approximately $45\%$. This indicates that nearly half of the selected variables are on average non-informative.

\begin{table}[!htpb]
\caption{\label{tab:sel}Average selection rates in the low-dimensional linear setting with 100 repetitions, the spatial error family, model-based gradient boosting with first step gradient boosting with deselection (DS-GB) across different spatial autoregressive parameters $\lambda$. Reported are the true positive rate (TPR), true negative rate (TNR), and false discovery rate (FDR).}
\centering
\begin{tabular}{@{\extracolsep{5pt}}cccc}
\\[-1.8ex] \hline
\hline \\[-1.8ex]
$\lambda$ & TPR & TNR & FDR \\
\hline \\[-1.8ex]
$-0.8$ & 100\% & 25.31\% & 74.59\% \\
$-0.6$ & 100\% & 26.63\% & 74.30\% \\
$-0.4$ & 100\% & 26.69\% & 74.25\% \\
$-0.2$ & 100\% & 25.69\% & 74.47\% \\
$0.2$  & 100\% & 30.75\% & 73.04\% \\
$0.4$  & 100\% & 34.63\% & 71.79\% \\
$0.6$  & 100\% & 50.94\% & 64.67\% \\
$0.8$  & 100\% & 75.50\% & 45.48\% \\
\hline \\[-1.8ex]
\end{tabular}
\end{table}

\begin{figure}[H] 
    \centering
    \includegraphics[width = \textwidth]{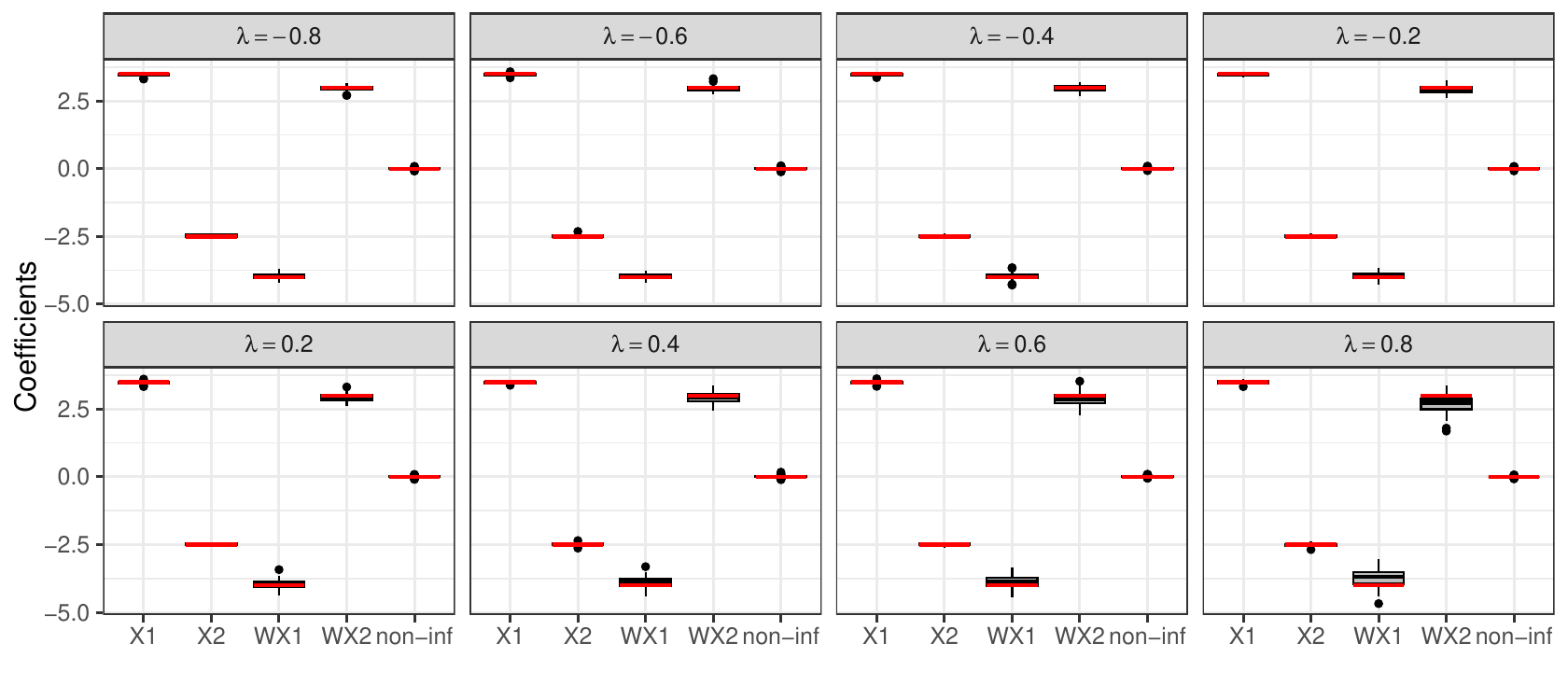}
    \caption{Estimated linear effects for the low-dimensional linear setting with 100 repetitions, the spatial error family, model-based gradient boosting with first step gradient boosting with deselection (DS-GB) across different spatial autoregressive parameters $\lambda$. Horizontal red lines represent the true values.}
    \label{fig:est1}
\end{figure}

The performance of estimation for the linear effects in low-dimensional linear setting can be seen in Figure \ref{fig:est1}. The estimated coefficients are presented by boxplots where the red horizontal line represents the true coefficient values according to the data generating process. Across all spatial autoregressive parameters $\lambda$, DS-GB manages to recover all informative variables. However, the coefficient estimates are slightly biased across all values of $\lambda$. In general, the magnitude of bias is stronger for spatial lags of the exogenous variables which is additionally increasing as $\lambda$ increases. However, a bias in the estimates for DS-GB is unsurprising and actually expected since the presented estimates for the linear effects are based on regularization via early stopping. In contrast, non-informative variables where rarely chosen with coefficient estimates clustered around $0$.

Additionally, the impact of the choice of the estimation method in the first step on the estimates of the $\lambda$ parameter are evaluated. The results are shown in Table \ref{tab:bias}. Across all different spatial autoregressive parameters $\lambda$, the classical estimation strategies, namely QML and GMM struggle to reliable estimate $\lambda$ indicated by substantial downward biases. However, the bias seems to decrease as $\lambda$ increases and becomes positive yielding best values for $\lambda = 0.8$. Furthermore, QML outperforms corresponding GMM for positive $\lambda$ values. The results for LS-GB are identical to GMM, since the first step is based on OLS in both estimation strategies. The results change substantially for GB-GB and DS-GB. In principle, utilizing model-based gradient boosting in the first step to obtain consistent estimates of $\bm{\delta}$ leads to a decrease in the bias, MSE and ESE. The decrease in bias, MSE and ESE is further improved if model-based gradient boosting is combined with deselection in the first step. Moreover, the results show that GB-GB and DS-GB consistently outperform corresponding QML and GMM estimators.

\begin{table}[!htpb]
\caption{\label{tab:bias}Estimation performance for the spatial autoregressive parameter $\lambda$ in the low-dimensional linear setting with 100 repetitions and the spatial error family. Reported are the bias, mean squared error (MSE) (in parentheses), and empirical standard error (ESE) [in brackets] for quasi-maximum likelihood (QML), generalized method of moments (GMM), and model-based gradient boosting with first step OLS (LS-GB), gradient boosting (GB-GB), and gradient boosting with deselection (DS-GB).}
\centering
\small
\begin{tabular}{@{\extracolsep{5pt}}cccccc}
\\[-1.8ex] \hline
\hline \\[-1.8ex]
$\lambda$ & QML & GMM & LS-GB & GB-GB & DS-GB \\
\hline \\[-1.8ex]

\multirow{3}{*}{$-0.8$} 
& -0.1298 & -0.0847 & -0.0847 & -0.0484 & -0.0337 \\
& (0.0461) & (0.0381) & (0.0381) & (0.0296) & (0.0313) \\
& [0.1718] & [0.1769] & [0.1769] & [0.1659] & [0.1746] \\

\multirow{3}{*}{$-0.6$} 
& -0.1525 & -0.1197 & -0.1197 & -0.0888 & -0.0530 \\
& (0.0483) & (0.0423) & (0.0423) & (0.0367) & (0.0297) \\
& [0.1590] & [0.1683] & [0.1683] & [0.1707] & [0.1647] \\

\multirow{3}{*}{$-0.4$} 
& -0.1559 & -0.1372 & -0.1372 & -0.0966 & -0.0536 \\
& (0.0452) & (0.0419) & (0.0419) & (0.0262) & (0.0169) \\
& [0.1454] & [0.1527] & [0.1527] & [0.1304] & [0.1191] \\

\multirow{3}{*}{$-0.2$} 
& -0.1408 & -0.1335 & -0.1335 & -0.1052 & -0.0661 \\
& (0.0485) & (0.0454) & (0.0454) & (0.0373) & (0.0303) \\
& [0.1701] & [0.1669] & [0.1669] & [0.1628] & [0.1618] \\

\multirow{3}{*}{$0.2$} 
& -0.1015 & -0.1166 & -0.1166 & -0.0873 & -0.0362 \\
& (0.0259) & (0.0280) & (0.0280) & (0.0209) & (0.0137) \\
& [0.1256] & [0.1208] & [0.1208] & [0.1157] & [0.1117] \\

\multirow{3}{*}{$0.4$} 
& -0.0862 & -0.1105 & -0.1105 & -0.0919 & -0.0403 \\
& (0.0175) & (0.0216) & (0.0216) & (0.0167) & (0.0084) \\
& [0.1010] & [0.0973] & [0.0973] & [0.0910] & [0.0827] \\

\multirow{3}{*}{$0.6$} 
& -0.0530 & -0.0896 & -0.0896 & -0.0772 & -0.0236 \\
& (0.0100) & (0.0154) & (0.0154) & (0.0127) & (0.0052) \\
& [0.0852] & [0.0865] & [0.0865] & [0.0826] & [0.0683] \\

\multirow{3}{*}{$0.8$} 
& -0.0168 & -0.0492 & -0.0492 & -0.0483 & -0.0147 \\
& (0.0023) & (0.0051) & (0.0051) & (0.0048) & (0.0018) \\
& [0.0453] & [0.0521] & [0.0521] & [0.0498] & [0.0398] \\

\hline \\[-1.8ex]
\end{tabular}
\end{table}

Finally, the prediction performance on an independent test data set for the low-dimensionsal linear setting is evaluated. The results can be seen in Table \ref{tab:pred_perf}. In principle, the RMSEP, MAEP, and NLL are consistently higher across all values of the spatial autoregressive parameter $\lambda$ for QML and GMM, compared to the model-based gradient boosting algorithms. In contrast, the predictive performance of the model-based gradient boosting algorithms remains stable, with only minor differences at the decimal level.

\begin{table}[!htpb]
\caption{\label{tab:pred_perf}Prediction performance on independent test data for the low-dimensional linear setting with 100 repetitions and the spatial error family across different spatial autoregressive parameters $\lambda$. Reported are the root mean squared error of prediction (RMSEP), mean absolute error of prediction (MAEP), and quasi negative log-likelihood (NLL) for quasi-maximum likelihood (QML), generalized method of moments (GMM), and model-based gradient boosting with first step OLS (LS-GB), gradient boosting (GB-GB), and gradient boosting with deselection (DS-GB).}
\centering
\begin{tabular}{@{\extracolsep{5pt}}clccccc}
\\[-1.8ex] \hline
\hline \\[-1.8ex]
$\lambda$ & Metric & QML & GMM & LS-GB & GB-GB & DS-GB \\
\hline \\[-1.8ex]
\multirow{3}{*}{$-0.8$}
& RMSEP & 1.0894 & 1.0894 & 1.0819 & 1.0819 & 1.0820 \\
& MAEP  & 0.8689 & 0.8689 & 0.8630 & 0.8629 & 0.8630 \\
& NLL   & 795.66 & 794.78 & 790.41 & 789.77 & 789.64 \\
\hline
\multirow{3}{*}{$-0.6$}
& RMSEP & 1.0509 & 1.0508 & 1.0440 & 1.0439 & 1.0438 \\
& MAEP  & 0.8393 & 0.8392 & 0.8339 & 0.8338 & 0.8337 \\
& NLL   & 786.96 & 786.24 & 782.17 & 781.51 & 780.79 \\
\hline
\multirow{3}{*}{$-0.4$}
& RMSEP & 1.0381 & 1.0381 & 1.0309 & 1.0309 & 1.0307 \\
& MAEP  & 0.8291 & 0.8291 & 0.8233 & 0.8234 & 0.8232 \\
& NLL   & 786.27 & 785.83 & 781.42 & 780.47 & 779.47 \\
\hline
\multirow{3}{*}{$-0.2$}
& RMSEP & 1.0326 & 1.0326 & 1.0234 & 1.0232 & 1.0229 \\
& MAEP  & 0.8243 & 0.8243 & 0.8168 & 0.8166 & 0.8163 \\
& NLL   & 785.93 & 785.72 & 780.80 & 780.06 & 779.18 \\
\hline
\multirow{3}{*}{$0.2$}
& RMSEP & 1.0360 & 1.0361 & 1.0234 & 1.0227 & 1.0220 \\
& MAEP  & 0.8293 & 0.8293 & 0.8192 & 0.8187 & 0.8181 \\
& NLL   & 782.44 & 782.76 & 777.51 & 776.67 & 775.57 \\
\hline
\multirow{3}{*}{$0.4$}
& RMSEP & 1.0831 & 1.0830 & 1.0656 & 1.0647 & 1.0631 \\
& MAEP  & 0.8650 & 0.8649 & 0.8511 & 0.8504 & 0.8493 \\
& NLL   & 788.86 & 789.48 & 783.58 & 782.69 & 781.10 \\
\hline
\multirow{3}{*}{$0.6$}
& RMSEP & 1.1651 & 1.1660 & 1.1416 & 1.1406 & 1.1368 \\
& MAEP  & 0.9308 & 0.9316 & 0.9120 & 0.9111 & 0.9078 \\
& NLL   & 788.89 & 790.00 & 784.18 & 783.47 & 781.31 \\
\hline
\multirow{3}{*}{$0.8$}
& RMSEP & 1.4697 & 1.4708 & 1.4397 & 1.4404 & 1.4400 \\
& MAEP  & 1.1732 & 1.1742 & 1.1476 & 1.1481 & 1.1483 \\
& NLL   & 803.06 & 803.62 & 797.28 & 797.14 & 796.00 \\
\hline \\[-1.8ex]
\end{tabular}
\end{table}

In general, the presented simulation results show proper functionality of the model-based gradient boosting algorithm from the perspective of specificity and sensitivity in the low-dimensional linear setting across all considered spatial autoregressive parameters $\lambda$. This conclusion can be drawn based on the average selection rates where true informative variables are consistently selected across all 100 repetitions on average. Furthermore, the estimated linear effects show correct direction and algebraic sign. Nevertheless, biases are introduced due to the inherent regularization of model-based gradient boosting via early stopping. Although the average selection rates show high TNR and FDR, the actual impact in terms of estimated coefficients for non-informative variables remains very low, meaning the included non-informative variables are entering the final model only with low values. To provide intuition regarding the consistency and convergence of the proposed model-based gradient boosting algorithm, results are presented through a comparison with GMM coefficient estimates. These can be recovered by utilizing a sufficiently large stopping criterion, as demonstrated in Appendix \ref{app:conv}. Additionally, model-based gradient boosting outperforms the established QML and GMM estimators in estimating $\lambda$ when substantial noise is introduced through non-informative variables. As a result, the QML and GMM estimate exhibits a downward bias in the presence of such noise. This outcome is not unexpected, as spatial autocorrelation typically decreases when informative variables are added to the model. However, the simulation study also indicates that introducing non-informative noise can falsely reduce spatial autocorrelation, which has serious implications for model interpretation as it may lead to highly misleading conclusions. Utilizing model-based gradient boosting can effectively mitigate the consequences by not including non-informative variables, thereby improving quality of residuals in the first step and decreasing the bias across all values of $\lambda$. While model-based gradient boosting consistently outperforms classical QML and GMM estimation strategies in terms of predictive accuracy on unobserved test data, a notable drawback is its relatively FDR, indicating that the final model may still include many non-informative variables. However, its ability to exclude such variables improves drastically as $\lambda$ increases. Despite this improvement, the persistently high FDR suggests that model complexity remains an issue, even under stronger spatial dependence.

\subsubsection{High-dimensional linear setting (n = 400, p = 800)}
Similar to the low-dimensional linear setting, results are also evaluated for the high-dimensional linear setting. Here, only the GB-GB and DS-GB approaches are reported, as model-based gradient boosting remains the only feasible method when the number of variables exceeds the number of observations. Table \ref{tab:sel2} presents the average selection rates. Across all values of the spatial autoregressive parameter $\lambda$, the TPR remains consistently at $100\%$, indicating that all informative variables are selected on average. In contrast to the low-dimensional linear setting, the TNR is also consistently high, demonstrating that most non-informative variables are successfully excluded. Notably, TNR improves with increasing $\lambda$, reaching approximately $94\%$ at $\lambda = 0.8$. Despite these promising findings, the FDR remains high suggesting that the final model still includes many non-informative variables. However, this outcome is not unexpected, given that only four informative variables are present and can easily be overshadowed by the large number of non-informative variables.

\begin{table}[!htpb]
\caption{\label{tab:sel2}Average selection rates in the high-dimensional linear setting with 100 repetitions, the spatial error family, model-based gradient boosting with first step gradient boosting with deselection (DS-GB) across different spatial autoregressive parameters $\lambda$. Reported are the true positive rate (TPR), true negative rate (TNR), and false discovery rate (FDR).}
\centering
\begin{tabular}{@{\extracolsep{5pt}}cccc}
\\[-1.8ex] \hline
\hline \\[-1.8ex]
$\lambda$ & TPR & TNR & FDR \\
\hline \\[-1.8ex]
$-0.8$ & 100\% & 86.84\% & 96.31\% \\
$-0.6$ & 100\% & 86.67\% & 96.36\% \\
$-0.4$ & 100\% & 86.79\% & 96.32\% \\
$-0.2$ & 100\% & 87.05\% & 96.25\% \\
$0.2$  & 100\% & 87.44\% & 96.14\% \\
$0.4$  & 100\% & 88.30\% & 95.85\% \\
$0.6$  & 100\% & 91.26\% & 94.40\% \\
$0.8$  & 100\% & 94.21\% & 91.65\% \\
\hline \\[-1.8ex]
\end{tabular}
\end{table}

\begin{figure}[!htpb] 
    \centering
    \includegraphics[width = \textwidth]{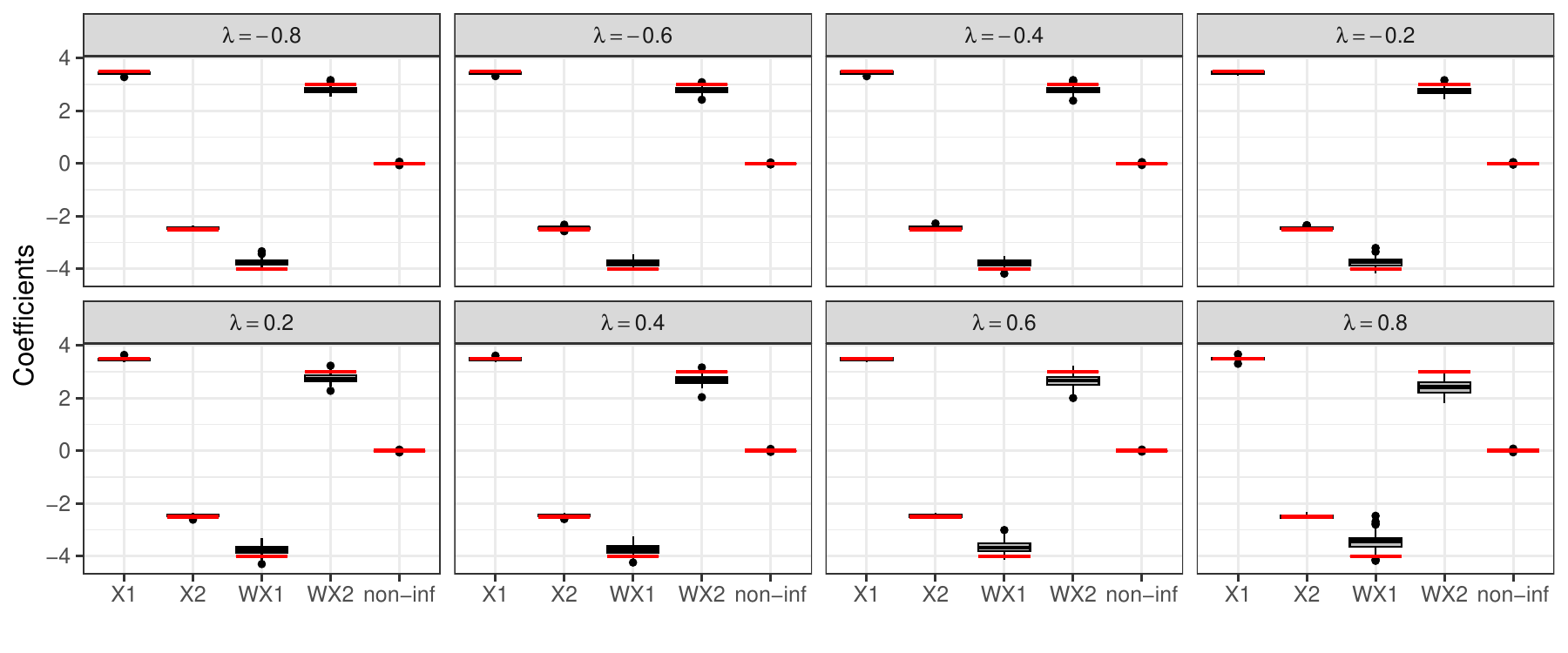}
    \caption{Estimated linear effects for the high-dimensional linear setting with 100 repetitions, the spatial error family, model-based gradient boosting with first step gradient boosting with deselection (DS-GB) across different spatial autoregressive parameters $\lambda$. Horizontal red lines represent the true values.}
    \label{fig:est2}
\end{figure}

The estimation performance of the coefficients in the high-dimensional linear setting is illustrated in Figure \ref{fig:est2}. Compared to the low-dimensional linear setting, the coefficient estimates exhibit a more pronounced shrinkage effect, particularly for the coefficients of the spatial lags of the exogenous variables. Furthermore, the shrinkage effect additionally intensifies as the spatial autoregressive parameter $\lambda$ increases and becomes more positive. Nevertheless, the algebraic signs and general direction of the coefficients remain consistently correct even in the presence of many non-informative variables. 

Furthermore, the estimation performance for the spatial autoregressive parameter $\lambda$ in the high-dimensional linear setting can be seen in Table \ref{tab:bias2}. In principle, QML, GMM and LS-GB are not feasible anymore. Thus only GB-GB and DS-GB remain as feasible estimation strategies. The results for GB-GB indicate an increasing bias on the estimated $\hat{\lambda}$. This behavior can be explained by the increased shrinkage effect on the informative variables in combination with many non-informative variables since the generalized method of moment estimator heavily relies on the quality of the predictions of the residuals. In contrast, DS-GB almost always deselect all non-informative variables. Thus, even though the shrinkage effect is present, the additional noise from non-informative variables is removed which ensures a proper recovery of the true $\lambda$.  

\begin{table}[!htpb]
\caption{\label{tab:bias2} Estimation performance for the spatial autoregressive parameter $\lambda$ in the high-dimensional linear setting with 100 repetitions and the spatial error family. Reported are the bias, mean squared error (MSE) (in parentheses), and empirical standard error (ESE) [in brackets] for model-based gradient boosting with gradient boosting (GB-GB) and gradient boosting with deselection (DS-GB).}
\centering
\small
\begin{tabular}{@{\extracolsep{5pt}}cccccc}
\\[-1.8ex] \hline
\hline \\[-1.8ex]
$\lambda$ & QML & GMM & LS-GB & GB-GB & DS-GB \\
\hline \\[-1.8ex]

\multirow{3}{*}{$-0.8$} 
& -- & -- & -- & 0.0351 & -0.0106 \\
& -- & -- & -- & (0.0196) & (0.0238) \\
& -- & -- & -- & [0.1360] & [0.1546] \\

\multirow{3}{*}{$-0.6$} 
& -- & -- & -- & -0.0232 & -0.0205 \\
& -- & -- & -- & (0.0184) & (0.0225) \\
& -- & -- & -- & [0.1343] & [0.1492] \\

\multirow{3}{*}{$-0.4$} 
& -- & -- & -- & -0.0658 & -0.0364 \\
& -- & -- & -- & (0.0156) & (0.0165) \\
& -- & -- & -- & [0.1065] & [0.1237] \\

\multirow{3}{*}{$-0.2$} 
& -- & -- & -- & -0.1273 & -0.0503 \\
& -- & -- & -- & (0.0266) & (0.0198) \\
& -- & -- & -- & [0.1025] & [0.1322] \\

\multirow{3}{*}{$0.2$} 
& -- & -- & -- & -0.2344 & -0.0415 \\
& -- & -- & -- & (0.0626) & (0.0125) \\
& -- & -- & -- & [0.0878] & [0.1041] \\

\multirow{3}{*}{$0.4$} 
& -- & -- & -- & -0.3251 & -0.0303 \\
& -- & -- & -- & (0.1111) & (0.0077) \\
& -- & -- & -- & [0.0738] & [0.0831] \\

\multirow{3}{*}{$0.6$} 
& -- & -- & -- & -0.4464 & -0.0321 \\
& -- & -- & -- & (0.2047) & (0.0049) \\
& -- & -- & -- & [0.0740] & [0.0621] \\

\multirow{3}{*}{$0.8$} 
& -- & -- & -- & -0.5883 & -0.0156 \\
& -- & -- & -- & (0.3515) & (0.0022) \\
& -- & -- & -- & [0.0737] & [0.0439] \\

\hline \\[-1.8ex]
\end{tabular}
\end{table}

Finally, the predictive performance on independent test data in the high-dimensional linear setting is shown in Table \ref{tab:pred_perf2}. As before, results are reported only for GB-GB and DS-GB, since QML, GMM, and LS-GB are infeasible in this setting. Although the performance itself is more varied, the absolute values of the performance criteria are comparable to those observed in the low-dimensional linear setting. As expected, the predictive accuracy of DS-GB improves with increasing values of $\lambda$ which corresponds to the reduction in bias observed for estimate $\hat{\lambda}$. However, a direct performance comparison with QML or GMM is not possible due to the infeasibility in the high-dimensional linear setting.

\begin{table}[!htpb]
\caption{\label{tab:pred_perf2}Prediction performance on independent test data for the high-dimensional linear setting with 100 repetitions and the spatial error family across different spatial autoregressive parameters $\lambda$. Reported are the root mean squared error of prediction (RMSEP), mean absolute error of prediction (MAEP), and quasi negative log-likelihood (NLL) for model-based gradient boosting with gradient boosting (GB-GB) and gradient boosting with deselection (DS-GB).}
\centering
\begin{tabular}{@{\extracolsep{5pt}}clccccc}
\\[-1.8ex] \hline
\hline \\[-1.8ex]
$\lambda$ & Metric & QML & GMM & LS-GB & GB-GB & DS-GB \\
\hline \\[-1.8ex]

\multirow{3}{*}{$-0.8$}
& RMSEP & -- & -- & -- & 1.1311 & 1.1318 \\
& MAEP  & -- & -- & -- & 0.9034 & 0.9039 \\
& NLL   & -- & -- & -- & 868.72 & 873.56 \\

\multirow{3}{*}{$-0.6$}
& RMSEP & -- & -- & -- & 1.1115 & 1.1126 \\
& MAEP  & -- & -- & -- & 0.8886 & 0.8895 \\
& NLL   & -- & -- & -- & 866.33 & 868.96 \\

\multirow{3}{*}{$-0.4$}
& RMSEP & -- & -- & -- & 1.0892 & 1.0908 \\
& MAEP  & -- & -- & -- & 0.8708 & 0.8720 \\
& NLL   & -- & -- & -- & 861.27 & 862.98 \\

\multirow{3}{*}{$-0.2$}
& RMSEP & -- & -- & -- & 1.0753 & 1.0775 \\
& MAEP  & -- & -- & -- & 0.8562 & 0.8580 \\
& NLL   & -- & -- & -- & 862.11 & 863.01 \\

\multirow{3}{*}{$0.2$}
& RMSEP & -- & -- & -- & 1.0828 & 1.0844 \\
& MAEP  & -- & -- & -- & 0.8627 & 0.8640 \\
& NLL   & -- & -- & -- & 862.92 & 853.88 \\

\multirow{3}{*}{$0.4$}
& RMSEP & -- & -- & -- & 1.1170 & 1.1111 \\
& MAEP  & -- & -- & -- & 0.8914 & 0.8870 \\
& NLL   & -- & -- & -- & 875.41 & 846.71 \\

\multirow{3}{*}{$0.6$}
& RMSEP & -- & -- & -- & 1.2187 & 1.1914 \\
& MAEP  & -- & -- & -- & 0.9731 & 0.9516 \\
& NLL   & -- & -- & -- & 913.89 & 829.33 \\

\multirow{3}{*}{$0.8$}
& RMSEP & -- & -- & -- & 1.5798 & 1.5134 \\
& MAEP  & -- & -- & -- & 1.2633 & 1.2130 \\
& NLL   & -- & -- & -- & 1066.07 & 825.05 \\

\hline \\[-1.8ex]
\end{tabular}
\end{table}

The results for the high-dimensional linear setting support the results of the low-dimensional linear settings and indicate proper functionality of the model-based gradient boosting algorithm from the perspective of specificity and sensitivity across all spatial autoregressive parameters $\lambda$. The TPR remains consistently at $100\%$ across all 100 repetitions on average. Although the estimated coefficients are stronger affected from shrinkage, the algebraic sign and general direction remain correct. The GB-GB exhibits a strong bias as $\lambda$ increases indicating that deselection in the first step is necessary to reliably estimate $\lambda$ in high-dimensional linear settings. Nevertheless, the great advantage is the feasibility of model-based gradient boosting in scenarios where the number of variables exceed the number of observations. In such situations, the established estimation strategies like QML and GMM fail entirely evidenced in the simulation study by missing values. For instance, the predictive performance as measured by the evaluation criteria, remains comparable in absolute terms to that of the low-dimensional linear setting. However, the results also reveal a consistently high FDR in the variable selection evaluation. Notably, no distinction is made between variables that are frequently selected and those that are selected only once, meaning all variables are equally weighted in the FDR computation. To address this limitation, the post-hoc deselection algorithm as proposed by \cite{stromer2022} is utilized to improve the FDR for model-based gradient boosting algorithm by removing variables that are likely selected by the algorithm merely by chance.

\subsubsection{Deselection}
As evidenced in the results for the low- as well as high-dimensional linear setting, model-based gradient boosting for SDEM suffers from a "greedy" selection behavior which attributes to the inclusion of too many non-informative variables, thereby decreasing TNR and increasing FDR. The reason for the "greedy" nature lies in the fact that the solutions of model-based gradient boosting are optimal with respect to the $L_1$-arc-length. In fact, by taking any convex loss function, model-based gradient boosting is able to approximate the solution path of a strictly monotone $L_1$-regularized regression model. Therefore, model-based gradient boosting is unable to automatically deselect a variable once it has been added to the model \citep{hastie2007, hepp2016}. To mitigate the consequences, the deselection algorithm for generalized linear, additive and interaction models is adapted to model-based gradient boosting and evaluated in the following \citep{stromer2022}. The average selection rates after utilizing model-based gradient boosting with first step gradient boosting with deselection and an additional deselection step (DS-DS) are presented in Table \ref{tab:des}.

\begin{table}[!htpb]
\caption{\label{tab:des} Average selection rates for model-based gradient boosting with first step gradient boosting with deselection and additional deselection (DS-DS) in the low- and high-dimensional linear setting with 100 repetitions, spatial error family across different spatial autoregressive parameters $\lambda$. Reported are the true positive rate (TPR), true negative rate (TNR) and false discovery rate (FDR).}
\centering
\begin{minipage}{0.48\textwidth}
\centering
\textbf{Low-Dimension}
\vspace{0.5em}
\begin{tabular}{@{\extracolsep{5pt}}cccc}
\\[-1.8ex] \hline
\hline \\[-1.8ex]
$\lambda$ & TPR & TNR & FDR \\
\hline \\[-1.8ex]
$-0.8$ & 100\% & 100\% & 0\% \\
$-0.6$ & 100\% & 100\% & 0\% \\
$-0.4$ & 100\% & 100\% & 0\% \\
$-0.2$ & 100\% & 100\% & 0\% \\
$0.2$  & 100\% & 100\% & 0\% \\
$0.4$  & 100\% & 100\% & 0\% \\
$0.6$  & 100\% & 100\% & 0\% \\
$0.8$  & 100\% & 100\% & 0\% \\
\hline \\[-1.8ex]
\end{tabular}
\end{minipage}
\hfill
\begin{minipage}{0.48\textwidth}
\centering
\textbf{High-Dimension}
\vspace{0.5em}
\begin{tabular}{@{\extracolsep{5pt}}cccc}
\\[-1.8ex] \hline
\hline \\[-1.8ex]
$\lambda$ & TPR & TNR & FDR \\
\hline \\[-1.8ex]
$-0.8$ & 100\% & 100\% & 0\% \\
$-0.6$ & 100\% & 100\% & 0\% \\
$-0.4$ & 100\% & 100\% & 0\% \\
$-0.2$ & 100\% & 100\% & 0\% \\
$0.2$  & 100\% & 100\% & 0\% \\
$0.4$  & 100\% & 100\% & 0\% \\
$0.6$  & 100\% & 100\% & 0\% \\
$0.8$  & 100\% & 100\% & 0\% \\
\hline \\[-1.8ex]
\end{tabular}
\end{minipage}
\end{table}

Indeed, the results clearly show that the TNR consistently remains at $100\%$ while the FDR decreases to $0\%$ on average. Thus, DS-DS successfully avoids selecting non-informative variables over all 100 repetitions, ensuring that only informative variables are included in the final model. This behavior is observed in low- as well as high-dimensional linear settings. Therefore, the deselection algorithm is able to mitigate the consequences of the inclusion of additional non-informative variables in boosted SDEM in both low- and high-dimensional linear settings. In fact, utilizing model-based gradient boosting in combination with deselection is strongly encouraged to ensure proper model and variable selection properties.

\section{Case study: Modeling life expectancy in German districts} \label{sec:case}
Life expectancy, defined as the average number of years a newborn in a given population is expected to live under current mortality conditions, is a statistical measure widely used to assess a variety of health, social, and economic outcomes \citep{marmot2005, cutler2006}. Over the past several decades, the life expectancy has been steadily increasing across Europe indicating progress in healthcare, socioeconomic development, and public health interventions. Notably, countries in the European Union report an increase in longevity due to improved living standards, better medical care, and lower mortality from major diseases such as cardiovascular conditions and cancer. Nevertheless, significant regional disparities remain between and within countries. For instance, Germany, one of the largest and economically wealthiest nation in the European Union, has seen signs of stagnation in life expectancy in comparison to especially Northern European countries. Thus, the stagnation gives rise to important concerns about regional inequalities, healthcare system efficiency, lifestyle-related risks, and demographic changes \citep{oecd2024}. In this case study, the goal is to model the life expectancy in German districts by utilizing spatial regression models with autoregressive disturbances to answer questions regarding underlying socio-economic and other determinants of life expectancy while incorporating district-level geographical disparities. In principle, the case study builds on recent previous research such as \cite{lampert2019, rau2020, siegel2022, jasilionis2023, marinetti2023, hoebel2025} which investigates current trends, socio-economic drivers of life expectancy, impact of the COVID-19 pandemic and patterns of mortality from both district-level and complete country perspectives.

To this end, a large-scale real-world data set, namely INKAR, is utilized. Managed by the Bundesinstitut für Bau-, Stadt- und Raumforschung, INKAR is an interactive (online) atlas about the living situation in Germany. It contains over 600 unique socio-demographic, socio-economic, and environmental indicators for distinct geographical locations allowing for evaluations of urban and rural disparities. In principle, INKAR is a panel data set relying on information starting from the year 1995. However, in this case study, the focus is on the cross-sectional life expectancy in German districts in the year 2019. Therefore, INKAR in the version of 2021 is utilized \citep{BBSR2024}. The INKAR data set in the current as well as the 2021 version with the codebook for all indicators is freely available at \url{https://www.inkar.de/}. 

To construct the data set suitable for this case study, the following preprocessing steps are necessary. First, the original number of districts available in INKAR version is 401. However, the data is merged with the German map data from the year 2024 available at \url{https://www.bkg.bund.de/}. The map data in the most recent version includes only 400 districts because the district "Eisenach" has been merged with a neighboring district in 2022 \citep{BKG2025}. To ensure comparability between both data sets, "Eisenach" is thus removed from the INKAR 2021 data resulting in 400 districts of interest. Second, a spatial weight matrix has to be generated based on the specific spatial configuration of the German districts. Therefore, the centroids of the spatial polygons based on the shape file layer from the German map data are computed. To keep a close connection to the simulation study in Section \ref{sec:sim}, a k-nearest neighbor structure is created utilizing $K = 10$, indicating that each district is assigned its ten geographically closest neighbors. Afterward, the resulting spatial weight matrix is row-normalized. Third, since the focus is primary on the variable selection evaluation, the indicators are allowed to be quite general. Thus, any domain knowledge is not imposed in the selection of variables. However, dubious indicators like the indicator number of a district as well as indicators describing similar phenomena are removed before model estimation. Furthermore, transformations for appropriate variables, that is, centering and scaling, are applied to ensure comparability across variables in terms of the scale.

\begin{figure}[!htpb] 
    \centering
    \includegraphics[width = \textwidth]{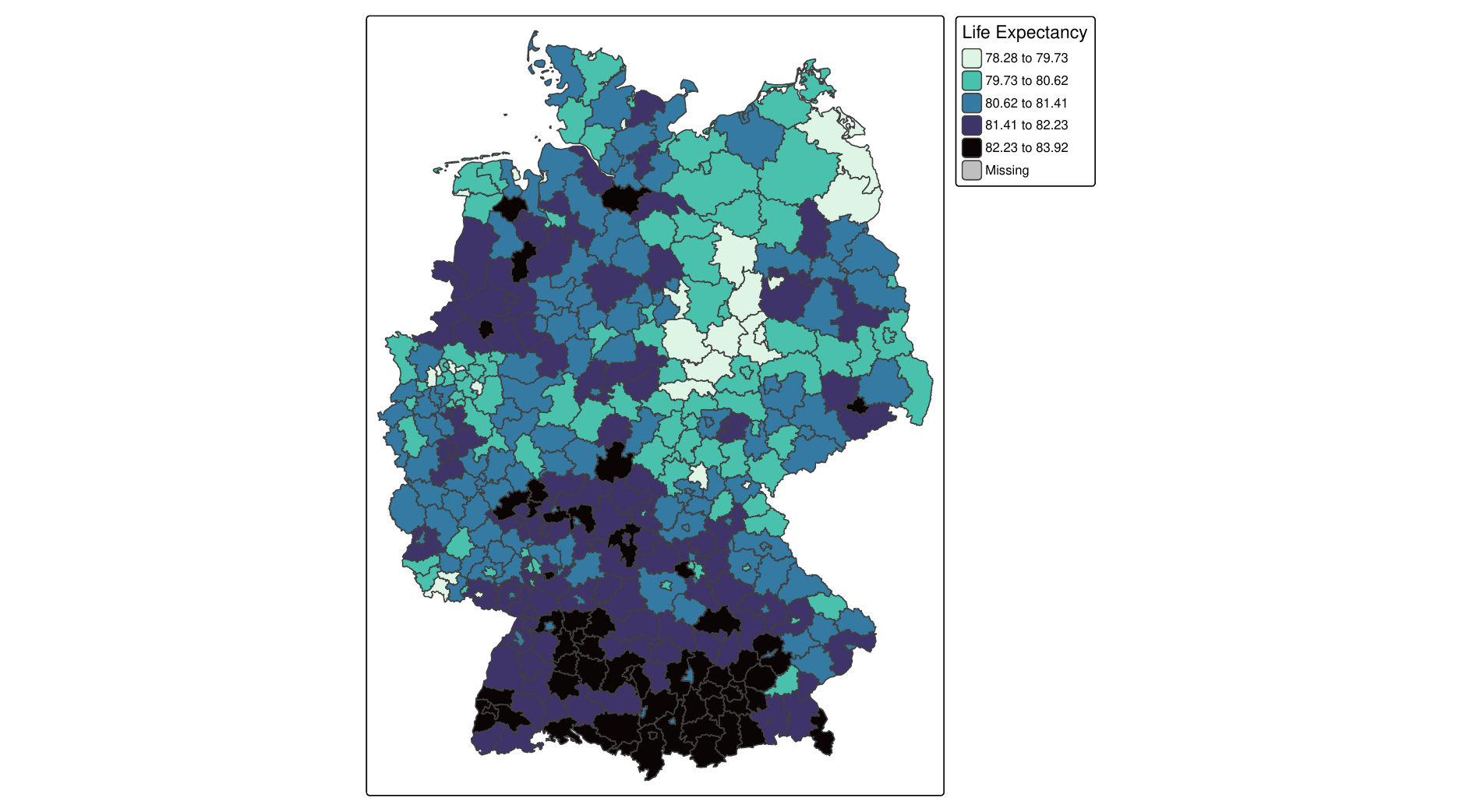}
    \caption{Life expectancy in German districts in the year 2019.}
    \label{fig:germany}
\end{figure}

The final data set is composed of 400 observations where each location corresponds to a district in Germany. These districts are divided again into 294 rural districts and 106 urban cities. A map of the life expectancy in German districts in the year 2019 can be seen in Figure \ref{fig:germany}.
The map reveals distinct disparities in life expectancy across German districts, with clear spatial clustering among neighboring regions. Notably, the highest life expectancy is observed in southern Bavaria, near the borders with Austria and Switzerland. Additional areas of high life expectancy can be found in Lower Saxony, close to the Dutch border. In contrast, life expectancy in the new federal states (former East Germany) is generally lower than in the old federal states (former West Germany), indicating a clear east-west divide. Details on the data types and descriptions of the dependent and independent variables of interest are provided in Table \ref{tab:ger}.
\begin{table}[!htpb]
\caption{Data type and description of dependent and independent variables for German district data.}
\label{tab:ger}
\begin{tabular}{@{\extracolsep{5pt}}llp{7.5cm}}
\\[-1.8ex] \hline
\hline \\[-1.8ex]
\textbf{Variable} & \textbf{Type} & \textbf{Description} \\ 
\hline \\[-1.8ex]
LIFE       & Numeric &  Average life expectancy of a newborn in years\\
AGE        & Numeric &  Average age of the population in years \\
UNEMPLOYMENT & Numeric & Percentage of unemployed among the civilian labor force \\
EMPLOYMENT & Numeric &  Percentage of social insurance-contributing employees per 100 working-age residents at place of residence \\
PART       & Numeric & Labor force participation rate among the working-age population in percent \\
SELF       & Numeric & Percentage of self-employed among all employed individuals \\
DEBT       & Numeric & Percentage of private debtors among residents aged 18 and over \\
WELFARE    & Numeric & Share of employable and non-employable welfare recipients among residents under 65 years old \\
ACADEMICS  & Numeric & Share of socially insured employees at the workplace with an academic professional qualification among socially insured employees in percent \\
MARRIAGES  & Numeric & Marriages per 1,000 inhabitants aged 18 and over \\
DIVORCES   & Numeric & Divorces per 1,000 inhabitants aged 18 and over \\
FOREIGN    & Numeric & Share of foreigners among the residents in percent \\
MEDINC     & Numeric & Median income of full-time employees subject to social security contributions in euro \\
HHINC      & Numeric & Average household income per inhabitant in euro \\
INS        & Numeric & Consumer insolvency proceedings per 1,000 inhabitants aged 18 and over \\
LABOR      & Numeric &  Number of hours worked by employees\\
LAND       & Numeric &  Cadastral area in square kilometers \\
LIVE       & Numeric &  Living space per inhabitant in square meters \\
URBAN      & Numeric &  Share of settlement and traffic area of the total area in percent \\
FOREST     & Numeric &  Share of forest area of the total area in percent \\
RECR       & Numeric &  Share of recreational area of the total area in percent\\
WATER      & Numeric &  Share of water area of the total area in percent \\
RENT       & Numeric &  Rent prices per square meter in euro \\
TAX        & Numeric &  Tax revenue per inhabitant in euro\\
GDP        & Numeric &  Gross domestic product per inhabitant \\
HOSP       & Numeric &  Medical care provision per 1,000 inhabitants\\
DR         & Numeric &  Medical doctors per 10,000 inhabitant \\
CARE       & Numeric &  Care-dependent individuals per 100 inhabitants \\
POP        & Numeric &  Inhabitant per square kilometer \\
CAR        & Numeric &  Passenger cars per 1,000 inhabitants \\
COM        & Numeric &  Commuter balance per 100 socially insured employees at the workplace \\
ACC        & Numeric &  Total road traffic accidents per 100,000 inhabitants \\
TRF        & Numeric &  Road traffic fatalities per 100,000 inhabitants \\
\hline
\end{tabular}
\end{table}

As evidenced by Figure \ref{fig:germany}, ignoring spatial dependence in the estimation process may lead to a misrepresentation of the underlying determinants influencing life expectancy. Therefore, the objective is to model life expectancy in German districts for the year 2019 using spatial regression models with autoregressive disturbances. In particular, the SDEM is employed, as spatial dependence may arise not only in the disturbances but also through spatial lags of the exogenous variables. For example, the number of medical doctors in neighboring districts may influence life expectancy, especially given the relatively short distances between locations. Generally, the results are presented with a clear focus on model-based gradient boosting. Since the application settings is a low-dimensional linear setting, results are reported for the QML, GMM, LS-GB, GB-GB, DS-GB and DS-DS defined exactly as in the simulation study in Section \ref{sec:sim}. All exogenous variables presented in Table \ref{tab:ger} as well as the corresponding first-order spatial lags and an intercept are included in the estimation yielding $65$ coefficients to estimate in a non-parsimonious model. Regarding the setup for the model-based gradient boosting algorithm, the narrative from the simulation study is followed and the learning rate is thus set to $s = 0.1$. Furthermore, for the optimization of the stopping criterion $m_{\text{stop}}$, the search is conducted by minimizing the empirical risk via 25-fold subsampling. For the deselection algorithm, $\tau$ is set to $0.01$. The estimated coefficients for all estimation strategies can be seen in Table \ref{tab:gerres}.

\begin{table}[!htpb]
\centering
\caption{\label{tab:gerres} Coefficient estimates in German district data for quasi-maximum likelihood (QML), generalized method of moments (GMM), model-based gradient boosting with first step OLS (LS-GB), gradient boosting (GB-GB), gradient boosting with deselection (DS-GB) and gradient boosting with deselection and additional deselection (DS-DS).}
\begin{tabular}{@{\extracolsep{5pt}}lcccccc}
\\[-1.8ex] \hline
\hline \\[-1.8ex]
{} & QML & GMM & LS-GB & GB-GB & DS-GB & DS-DS \\
\hline \\[-1.8ex]
$\lambda$         & 0.3887 & 0.2522 & 0.2522 & 0.4718 & 0.6430 &  0.6430      \\
$\text{Intercept}$& 81.1748 & 81.1771 & 81.1684 & 81.1621 & 81.1506 & 81.1506 \\
AGE               & -0.0763 & -0.0764 & -0.0783 & -0.0550 & -0.0355 & -0.0339 \\
UNEMPLOYMENT      & -0.3503 & -0.3401 & -0.2758 & -0.2651 & -0.1934 & -0.2552 \\
EMPLOYMENT        & -0.0165 & -0.0102 & -0.0018 & -0.0149 &         &         \\
PART              & 0.0319  & 0.0332  & 0.0056  &         &         &         \\
SELF              & 0.1162  & 0.1226  & 0.0904  & 0.0552  & 0.0417  &         \\
DEBT              & -0.3923 & -0.3819 & -0.3556 & -0.3602 & -0.3910 & -0.3910 \\
WELFARE           & 0.0974  & 0.0841  & 0.0722  &         &         &         \\
ACADEMICS         & 0.2326  & 0.2360  & 0.2016  & 0.1815  & 0.1281  & 0.0785  \\
MARRIAGES         & -0.0040 & -0.0048 & -0.0058 & -0.0022 &         &         \\
DIVORCES          & 0.0261  & 0.0241  & 0.0308  & 0.0365  & 0.0278  &         \\
FOREIGN           & 0.1501  & 0.1454  & 0.0832  & 0.0722  &         &         \\
MEDINC            & 0.0501  & 0.0522  & 0.0268  &         &         &         \\
HHINC             & 0.0723  & 0.0723  & 0.0583  & 0.0481  & 0.0326  & 0.0506  \\
INS               & -0.0104 & -0.0108 & -0.0079 & -0.0068 &         &         \\
LABOR             & -0.0258 & -0.0245 & -0.0501 & -0.0442 & -0.0293 &         \\
LAND              & -0.0368 & -0.0373 & -0.0483 & -0.0434 & -0.0132 &         \\
LIVE              & -0.0035 & -0.0019 &         &         &         &         \\
URBAN             & 0.1159  & 0.1274  &         &         &         &         \\
FOREST            & 0.0439  & 0.0448  & 0.0252  & 0.0186  &         &         \\
RECR              & 0.1191  & 0.1187  & 0.0688  & 0.0364  &         &         \\
WATER             & -0.0126 & -0.0144 & -0.0042 & -0.0699 &         &         \\
RENT              & 0.1598  & 0.1648  & 0.2001  & 0.2332  & 0.1985  & 0.2109  \\
TAX               & -0.0686 & -0.0724 & -0.0454 & -0.0347 &         &         \\
GDP               & -0.1146 & -0.1170 & -0.0750 & -0.0473 & -0.0092 &         \\
HOSP              & -0.0720 & -0.0730 & -0.0702 & -0.0549 & -0.0027 &         \\
DR                & 0.0551  & 0.0576  & 0.0345  & 0.0071  &         &         \\
CARE              & -0.1111 & -0.1123 & -0.0878 & -0.0953 & -0.1189 & -0.1324 \\
POP               & -0.2568 & -0.2661 & -0.1086 & -0.0808 &         &         \\
CAR               & 0.0371  & 0.0360  & 0.0054  & -0.1009 &         &         \\
COM               & -0.0522 & -0.0466 & -0.0407 & -0.0533 & -0.0539 &         \\
ACC               & 0.0528  & 0.0565  & 0.0408  & 0.0206  &         &         \\
TRF               & -0.0335 & -0.0343 & -0.0183 & -0.0161 & -0.0082 &         \\
$\sigma$          &  0.3452 & 0.3482  & 0.3610  & 0.3629  & 0.3812  &  0.3978   \\
\hline \\[-1.8ex]
$\text{No. of } \bm{W}$ & 32 & 32 & 23 & 15 & 0 & 0 \\
\hline \\[-1.8ex]
\end{tabular}
\end{table}

For brevity, the coefficients of the spatial lags of exogenous variables are not reported individually. Instead, the number of spatial lag variables (No. of $\bm{W}$) included in each model is presented. Regarding variable selection, the LS-GB approach reduces the total number of coefficients from $65$ to $53$ out of which $23$ can be attributed to the spatial lags of exogenous variables. A similar but improved outcome is observed for GB-GB where the number of variables is reduced to $41$ out of which $15$ correspond to the coefficients of the spatial lags of exogenous variables. The  best results in terms of variables selection are achieved when model-based gradient boosting is combined with deselection. For DS-GB, the number of selected variables is reduced to $18$ with no coefficients of the spatial lags of exogenous variables included. DS-DS further improves performance of variable selection through post-hoc deselection, resulting in a final model with only $8$ variables. These results suggest that many of the variables listed in Table \ref{tab:ger} are non-informative. Notably, model-based gradient boosting combined with deselection effectively reduces the initially non-parsimonious SDEM to a simpler and more parsimonious SEM in a data-driven manner, without the need for formal likelihood-based model selection criteria.

Regarding the spatial autoregressive parameter, all estimation strategies consistently indicate positive spatial dependence between neighboring districts, although the strength varies. Consistent with the downward bias observed in the simulation study, both QML and GMM estimators tend to underestimate the spatial autoregressive parameter $\lambda$. This underestimation can be attributed to the inclusion of non-informative variables which are largely eliminated through model-based gradient boosting. Consequently, estimates of $\lambda$ obtained from model-based gradient boosting are generally higher, depending on the first step estimation method. Moreover, the algebraic sign and overall direction of the coefficients remain largely consistent across all estimation strategies. The primary differences lie in the magnitude of the coefficients, which varies due to the regularization effects introduced by early stopping in model-based gradient boosting. Based on the results for DS-DS, the most important variables explaining the life expectancy in German districts in 2019 are the average age of inhabitants, the unemployment rate, the debt quota, the proportion of employees with an academic degree, the household income per inhabitant, the rent prices and the number of care-dependent individuals. Since the independent variables are transformed by simple scaling and centering, the coefficients have an intuitive and simple interpretation. For instance, an increase in rent prices by one euro ceteris paribus increases the average life expectancy by $0.2109$ years on average. Conversely, holding all other variables constant, a $1$ percentage point increase in the share of private debtors is associated with an average $0.3923$ year decrease in average life expectancy. Additionally, $\sigma$ increases when model-based gradient boosting is utilized which is unsurprising, as fewer variables are retained in the final model, potentially leaving more unexplained variation in the innovations.

To summarize the findings, the positive spatial autoregressive parameter $\lambda$ indicates strong spatial dependence between districts, that is, the life expectancy in one district is strongly influenced by the life expectancy in neighboring districts. Thus, health outcomes cluster spatially as suggested in Figure \ref{fig:germany}. Furthermore, an older average population is associated with lower life expectancy. Similarly, the number of care-dependent individuals negatively influence average life expectancy meaning that underlying poor health status decreases the average life expectancy. Districts with higher unemployment and debt quotas have on average lower average life expectancies indicating that financial hardship, fiscal stress or economic precarity is negatively correlated with longevity. In contrast, higher rent prices, higher household income and higher share of employees with an academic degree lead to a higher average life expectancy which shows that inhabitants in wealthier districts and better socioeconomic status live longer on average.

\section{Conclusion} \label{sec:con}
The key findings and main contributions of this article are: (a) Model-based gradient boosting is extended for spatial regression model with autoregressive disturbances, namely for the SDEM which unifies simpler models by considering spatial dependence in the linear predictor as well as the disturbances. (b) The algorithm is implemented in the \textbf{mboost} package \citep{bühlmann2007, hothorn2010, hofner2014, hofner2015} by providing a novel spatial error family. (c) Model-based gradient boosting for SDEM heavily relies on the knowledge about the spatial autoregressive parameter $\lambda$. However, in real-world application settings $\lambda$ is generally unknown making model-based gradient boosting infeasible. Therefore, a so-called feasible model-based gradient boosting algorithm for the SDEM is proposed which relies on replacing the unknown $\lambda$ in the ingredients by the corresponding estimated counterpart $\hat{\lambda}$ based on a generalized moment estimator proposed in \cite{kelejian1999}. (d) In extensive simulation studies for low- as well as high-dimensional linear settings, the results show proper functionality of the proposed feasible model-based gradient boosting algorithm. Particularly, estimation is accompanied by high TPR values and coefficients are estimated with high accuracy although biases are introduced due to regularization via early stopping. Model-based gradient boosting also outperforms standard QML and GMM estimators in terms of bias, MSE and ESE when estimating the autoregressive parameter $\lambda$ in the presence of non-informative variables. Additionally, the predictive performance is always better for model-based gradient boosting in comparison to QML and GMM. (e) The great advantage of model-based gradient boosting is the feasibility in high-dimensional linear settings where the number of variables exceeds the number of observations. Due to the modular nature, model-based gradient boosting does not require refitting and allows for a direct interpretation of the effects of coefficients on the dependent variable of interest enabling a potential application for a wide range of real-world settings. (f) Additionally, the feasible model-based gradient boosting is applied in a real-world application setting where the life expectancy in German districts is modeled. Additional case studies revisiting classical spatial econometric data sets, namely Boston housing prices \citep{harrison1978} and Columbus crime rate \citep{anselin1988} are provided in Appendix \ref{app:case}.

Naturally, limitations, improvements and extensions beyond the scope of this article have to be acknowledged. Although simulation studies have been provided, the scope of our simulation settings is of course by no means exhaustive. Further simulation studies for varying spatial weight matrices aim at partially address these limitations in Appendix \ref{app:mat} but much more complex scenarios can be easily considered. For instance, additional correlated noise variables or higher spatial lags of exogenous variables may be included in the simulation settings. A major limitation of the proposed model-based gradient boosting algorithm is the restricted applicability. Specifically, the algorithm can only be applied in settings where the dependent variable does not appear as a spatial lag in the spatial regression model. Thus, the implemented model-based gradient boosting algorithm is not readily extendable to spatial autoregressive models which are promising candidates for a wide range of applications. Additionally, the algorithm assumes homoskedastic innovations which is a potentially unrealistic assumption in many real-world application settings. In fact, heteroskedasticity is a common feature of spatial datasets \citep{lesage2009}. Revisiting the average selection rates reveals high FDR values indicating the inclusion of many non-informative variables of minor importance in the final model. Too mitigate the consequences, the utilization of a deselection algorithm is proposed. Alternatively, the application of the so-called stability selection could be utilized which enables the control over the amount of false positive \citep{meinshausen2010, shah2013, hofner2015}. Furthermore, probing can adapted for the model-based gradient boosting for the SDEM which is currently available for univariate location models. The general idea is to obtain sparser models by stopping the algorithm as soon as the first randomly permuted version of a variable is added \citep{thomas2017}. Finally, p-values for individual base-learners could be obtained by utilizing permutation techniques \citep{hepp2019}. 

Therefore, the next goal is to extend model-based gradient boosting for the SDEM by incorporating heteroskedastic innovations and adapting it to panel data models with spatially correlated error terms \citep{kapoor2007}. Such an extension would enable a more nuanced analysis of life expectancy across German districts. For example, the INKAR dataset provides panel data making it possible to study the underlying determinants of life expectancy over time. Beyond the spatial econometrics context, the novel model-based gradient boosting could be extended and applied to network econometric models \citep{lee2010}. Therefore, practitioners and applied statisticians working with spatial data are encouraged to utilize model-based gradient boosting for the SDEM as a valuable alternative for estimation, regularization, model and variable selection in future research.

\backmatter

\bmhead{Supplementary information}
The data as well as the accompanying codebook on the life expectancy in German districts is publicly available via \url{https://www.inkar.de/} (DL-DE BY 2.0).
All R-code for the implemented model-based gradient boosting algorithm along with the simulation studies is publicly available in the following GitHub repository \url{https://github.com/micbalz/SpatRegBoost}. 

\bmhead{Acknowledgments}
The work on this article was supported by the Deutsche Forschungsgemeinschaft (DFG, German Research Foundation) within project 492988838.

\section*{Declarations}

\bmhead{Conflict of interests}
The author declares that there is no conflict of interest.

\begin{appendices}
\section{Convergence of gradient boosting to generalized least squares estimates} \label{app:conv}

\begin{table}[H]
\centering
\caption{Comparison of coefficients estimated via generalized method of moments (GMM) and model-based gradient boosting with first step ordinary least squares (LS-GB) for a sufficiently high stopping criterion $m_{\text{stop}} = 25000$ and $\lambda = 0$.}
\begin{tabular}{@{\extracolsep{5pt}}lcc}
\\[-1.8ex] \hline
\hline \\[-1.8ex]
{} & \textbf{GMM} & \textbf{LS-GB} \\
\hline \\[-1.8ex]
$\lambda$         & 0.023925501 & 0.023925501 \\
$\bm{\text{Intercept}}$ & 0.965038052 & 0.965038052 \\
$\bm{X}_1$         & 3.556117719 & 3.556117719 \\
$\bm{X}_2$         & -2.390188072 & -2.390188072 \\
$\bm{X}_3$         & -0.097085565 & -0.097085565 \\
$\bm{X}_4$         & 0.029912130 & 0.029912130 \\
$\bm{X}_5$         & -0.020552856 & -0.020552856 \\
$\bm{X}_6$         & -0.002898073 & -0.002898073 \\
$\bm{X}_7$         & 0.011097265 & 0.011097265 \\
$\bm{X}_8$         & 0.044175354 & 0.044175354 \\
$\bm{X}_9$         & 0.009312604 & 0.009312604 \\
$\bm{X}_{10}$      & 0.039762952 & 0.039762952 \\
$\bm{WX}_1$        & -3.987535926 & -3.987535926 \\
$\bm{WX}_2$        & 2.902512886 & 2.902512886 \\
$\bm{WX}_3$        & -0.078164884 & -0.078164884 \\
$\bm{WX}_4$        & -0.121951105 & -0.121951105 \\
$\bm{WX}_5$        & -0.262497240 & -0.262497240 \\
$\bm{WX}_6$        & -0.129805748 & -0.129805748 \\
$\bm{WX}_7$        & -0.132761702 & -0.132761702 \\
$\bm{WX}_8$        & -0.011384246 & -0.011384246 \\
$\bm{WX}_9$        & -0.186093637 & -0.186093637 \\
$\bm{WX}_{10}$     & 0.119058680 & 0.119058680 \\
$\sigma$           & 1.006022029 & 1.006022029 \\
\hline \\[-1.8ex]
\end{tabular}
\label{tab:comparison}
\end{table}

\section{Further simulation results for varying spatial weight matrices} \label{app:mat}

\subsection{Low-Dimension}

\begin{table}[H]
\caption{Average selection rates in the low-dimensional linear setting with 100 repetitions, the spatial error family, model-based gradient boosting with first step gradient boosting with deselection (DS-GB) across different number of locations $K$. Reported are the true positive rate (TPR), true negative rate (TNR), and false discovery rate (FDR).}
\centering
\begin{tabular}{@{\extracolsep{5pt}}cccc}
\\[-1.8ex] \hline
\hline \\[-1.8ex]
$K$ & TPR & TNR & FDR \\
\hline \\[-1.8ex]
$1$ & 100\% & 28.13\% & 73.80\% \\
$2$ & 100\% & 29.75\% & 73.34\% \\
$3$ & 100\% & 28.13\% & 73.78\% \\
$5$ & 100\% & 27.44\% & 73.97\% \\
$10$ & 100\% & 28.19\% & 73.67\% \\
$20$ & 100\% & 44.50\% & 66.36\% \\
\hline \\[-1.8ex]
\end{tabular}
\end{table}

\begin{figure}[H] 
    \centering
    \includegraphics[width = \textwidth]{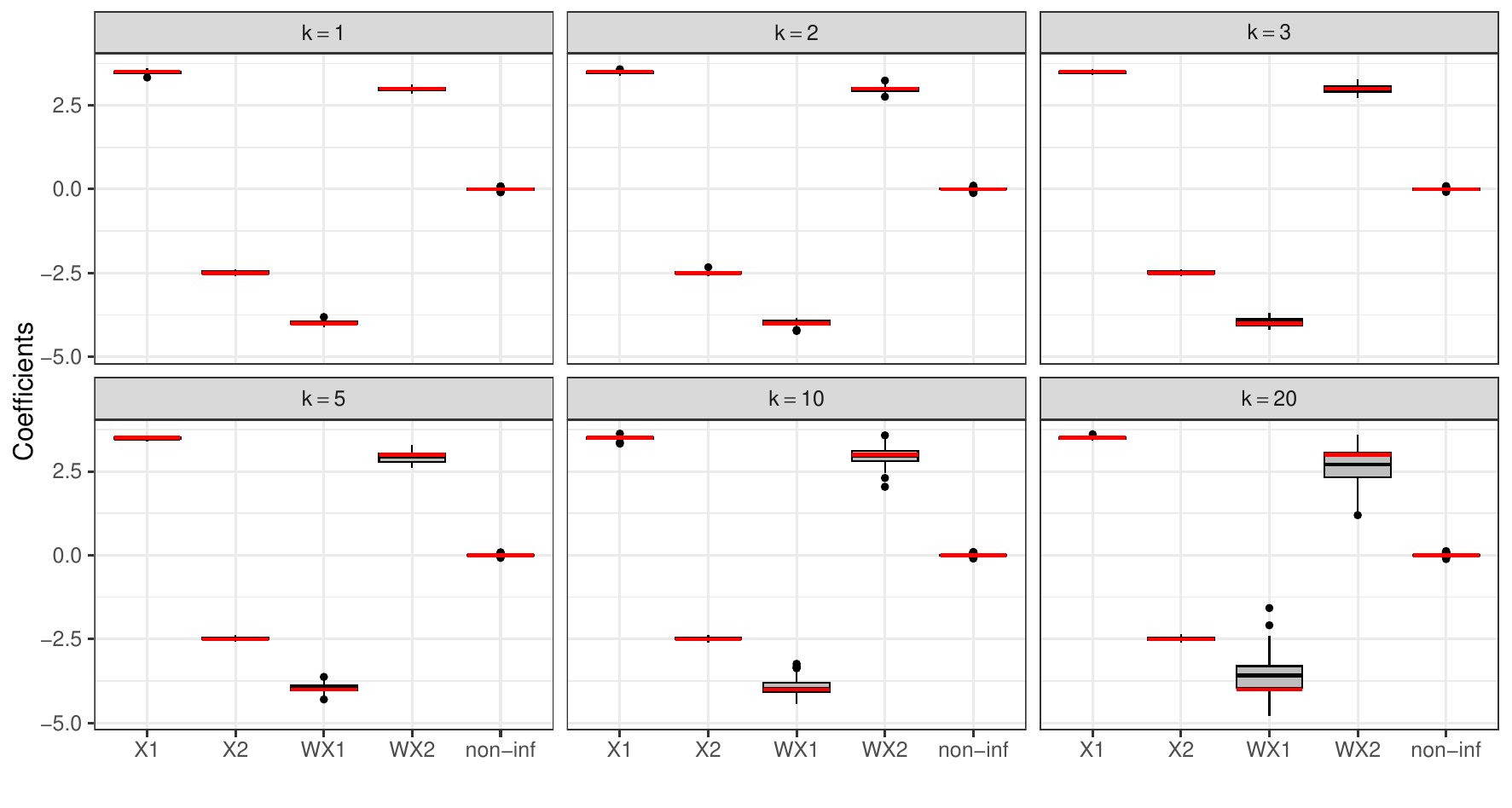}
    \caption{Estimated linear effects for the low-dimensional linear setting with 100 repetitions, the spatial error family, model-based gradient boosting with first step gradient boosting with deselection (DS-GB) across different number of locations $K$. Horizontal red lines represent the true values.}
\end{figure}

\begin{table}[H]
\caption{Estimation performance for the spatial autoregressive parameter $\lambda = 0$ in the low-dimensional linear setting with 100 repetitions and the spatial error family across different number of locations $K$. Reported are the bias, mean squared error (MSE) (in parentheses), and empirical standard error (ESE) [in brackets] for QML, GMM, LS-GB, model-based gradient boosting with gradient boosting (GB-GB), and gradient boosting with deselection (DS-GB).}
\centering
\small
\begin{tabular}{@{\extracolsep{5pt}}cccccc}
\\[-1.8ex] \hline
\hline \\[-1.8ex]
$K$ & QML & GMM & LS-GB & GB-GB & DS-GB \\
\hline \\[-1.8ex]

\multirow{3}{*}{$1$}
& -0.0049 & -0.0047 & -0.0047 & -0.0086 & -0.0062 \\
& (0.0024) & (0.0023) & (0.0023) & (0.0023) & (0.0021) \\
& [0.0491] & [0.0482] & [0.0482] & [0.0472] & [0.0459] \\

\multirow{3}{*}{$2$}
& -0.0259 & -0.0300 & -0.0300 & -0.0259 & -0.0114 \\
& (0.0063) & (0.0067) & (0.0067) & (0.0062) & (0.0054) \\
& [0.0756] & [0.0765] & [0.0765] & [0.0745] & [0.0727] \\

\multirow{3}{*}{$3$}
& -0.0625 & -0.0684 & -0.0684 & -0.0554 & -0.0282 \\
& (0.0141) & (0.0139) & (0.0139) & (0.0111) & (0.0075) \\
& [0.1014] & [0.0964] & [0.0964] & [0.0898] & [0.0823] \\

\multirow{3}{*}{$5$}
& -0.1294 & -0.1322 & -0.1322 & -0.1054 & -0.0620 \\
& (0.0415) & (0.0412) & (0.0412) & (0.0333) & (0.0251) \\
& [0.1583] & [0.1547] & [0.1547] & [0.1496] & [0.1466] \\

\multirow{3}{*}{$10$}
& -0.3158 & -0.3220 & -0.3220 & -0.2013 & -0.0794 \\
& (0.1718) & (0.1700) & (0.1700) & (0.0871) & (0.0542) \\
& [0.2697] & [0.2587] & [0.2587] & [0.2170] & [0.2200] \\

\multirow{3}{*}{$20$}
& -1.0065 & -0.9245 & -0.9245 & -0.4862 & 0.2400 \\
& (1.2601) & (1.1296) & (1.1296) & (0.3351) & (0.2648) \\
& [0.4995] & [0.5268] & [0.5268] & [0.3157] & [0.4575] \\

\hline \\[-1.8ex]
\end{tabular}
\end{table}

\begin{table}[H]
\caption{Prediction performance on independent test data for the low-dimensional linear setting with 100 repetitions and the spatial error family across different number of locations $K$. Reported are the root mean squared error of prediction (RMSE), mean absolute error of prediction (MAE), and quasi negative log-likelihood (NLL) for QML, GMM, LS-GB, model-based gradient boosting with gradient boosting (GB-GB), and gradient boosting with deselection (DS-GB).}
\centering
\small
\begin{tabular}{@{\extracolsep{5pt}}clccccc}
\\[-1.8ex] \hline
\hline \\[-1.8ex]
$K$ & Metric & QML & GMM & LS-GB & GB-GB & DS-GB \\
\hline \\[-1.8ex]

\multirow{3}{*}{$1$}
& RMSE & 1.0288 & 1.0288 & 1.0195 & 1.0193 & 1.0192 \\
& MAE  & 0.8192 & 0.8192 & 0.8120 & 0.8119 & 0.8118 \\
& NLL  & 781.87 & 781.85 & 777.66 & 777.61 & 777.45 \\

\multirow{3}{*}{$2$}
& RMSE & 1.0172 & 1.0172 & 1.0091 & 1.0089 & 1.0087 \\
& MAE  & 0.8107 & 0.8107 & 0.8047 & 0.8046 & 0.8044 \\
& NLL  & 776.88 & 776.96 & 773.34 & 773.17 & 772.90 \\

\multirow{3}{*}{$3$}
& RMSE & 1.0257 & 1.0257 & 1.0168 & 1.0166 & 1.0163 \\
& MAE  & 0.8202 & 0.8202 & 0.8130 & 0.8129 & 0.8126 \\
& NLL  & 781.19 & 781.23 & 776.98 & 776.62 & 775.97 \\

\multirow{3}{*}{$5$}
& RMSE & 1.0310 & 1.0309 & 1.0205 & 1.0202 & 1.0196 \\
& MAE  & 0.8229 & 0.8229 & 0.8144 & 0.8142 & 0.8137 \\
& NLL  & 784.98 & 785.03 & 779.99 & 779.19 & 778.12 \\

\multirow{3}{*}{$10$}
& RMSE & 1.0280 & 1.0278 & 1.0168 & 1.0158 & 1.0151 \\
& MAE  & 0.8234 & 0.8233 & 0.8138 & 0.8129 & 0.8120 \\
& NLL  & 789.75 & 789.79 & 783.43 & 779.51 & 777.02 \\

\multirow{3}{*}{$20$}
& RMSE & 1.0540 & 1.0524 & 1.0309 & 1.0284 & 1.0272 \\
& MAE  & 0.8428 & 0.8411 & 0.8245 & 0.8223 & 0.8215 \\
& NLL  & 857.56 & 841.98 & 817.40 & 792.66 & 780.40 \\

\hline \\[-1.8ex]
\end{tabular}
\end{table}

\subsection{High-Dimension}

\begin{table}[H]
\caption{Average selection rates in the high-dimensional linear setting with 100 repetitions, the spatial error family, model-based gradient boosting with first step gradient boosting with deselection (DS-GB) across different number of locations $K$. Reported are the true positive rate (TPR), true negative rate (TNR), and false discovery rate (FDR).}
\centering
\begin{tabular}{@{\extracolsep{5pt}}cccc}
\\[-1.8ex] \hline
\hline \\[-1.8ex]
$K$ & TPR & TNR & FDR \\
\hline \\[-1.8ex]
$1$  & 100\% & 86.86\% & 96.30\% \\
$2$  & 100\% & 86.70\% & 96.35\% \\
$3$  & 100\% & 86.91\% & 96.29\% \\
$5$  & 100\% & 87.20\% & 96.21\% \\
$10$ & 100\% & 87.52\% & 96.07\% \\
$20$ & 99\%  & 91.11\% & 93.67\% \\
\hline \\[-1.8ex]
\end{tabular}
\end{table}

\begin{figure}[H] 
    \centering
    \includegraphics[width = \textwidth]{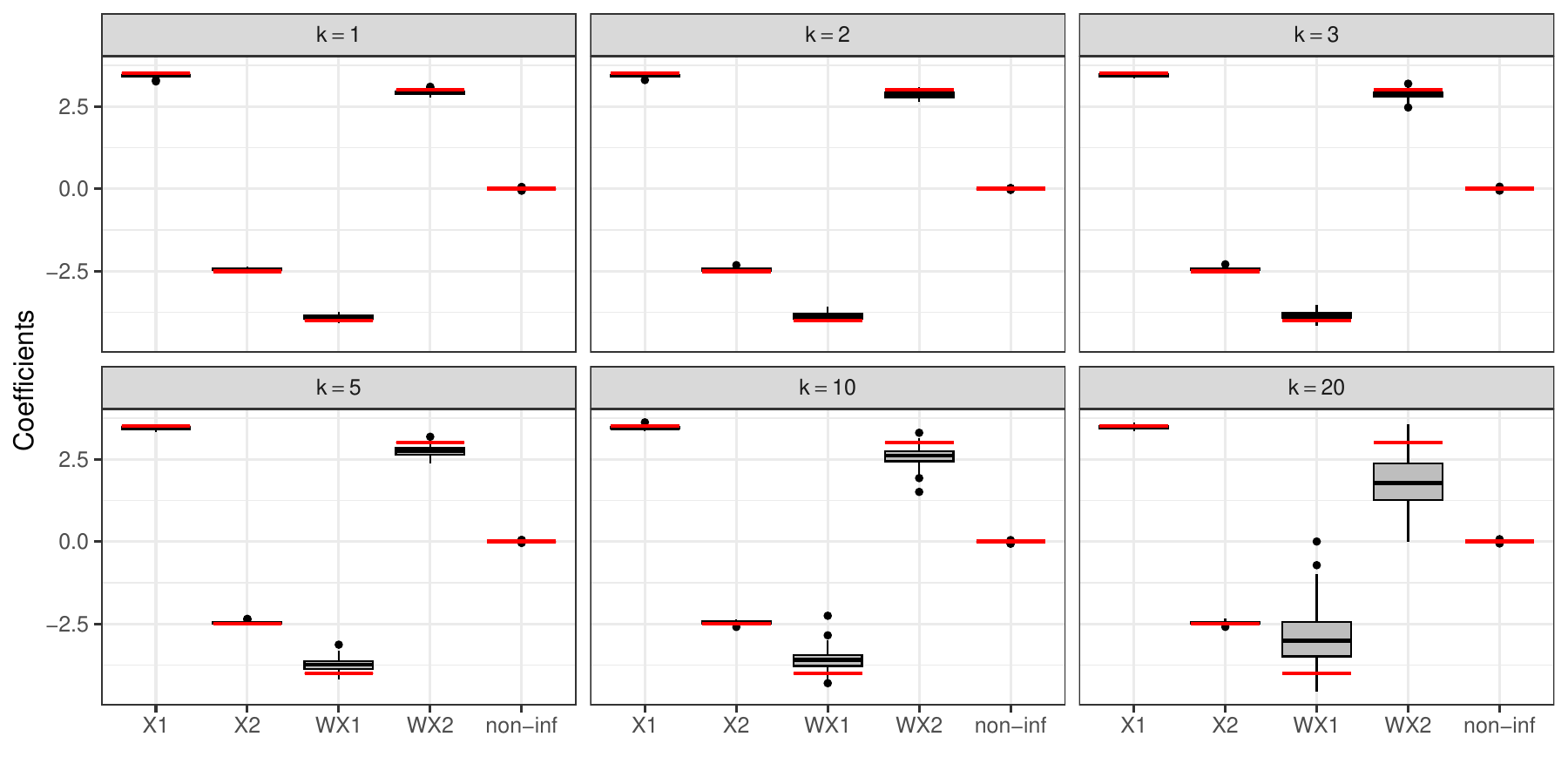}
    \caption{Estimated linear effects for the high-dimensional linear setting with 100 repetitions, the spatial error family, model-based gradient boosting with first step gradient boosting with deselection (DS-GB) across different number of locations $K$. Horizontal red lines represent the true values.}
\end{figure}

\begin{table}[H]
\caption{Estimation performance for the spatial autoregressive parameter $\lambda = 0$ in the high-dimensional linear setting with 100 repetitions and the spatial error family across different number of locations $K$. Reported are the bias, mean squared error (MSE) (in parentheses), and empirical standard error (ESE) [in brackets] for model-based gradient boosting with gradient boosting (GB-GB) and gradient boosting with deselection (DS-GB).}
\centering
\small
\begin{tabular}{@{\extracolsep{5pt}}cccccc}
\\[-1.8ex] \hline
\hline \\[-1.8ex]
$K$ & QML & GMM & LS-GB & GB-GB & DS-GB \\
\hline \\[-1.8ex]

\multirow{3}{*}{$1$} 
& -- & -- & -- & -0.0392 & -0.0114 \\
& -- & -- & -- & (0.0037) & (0.0024) \\
& -- & -- & -- & [0.0471] & [0.0483] \\

\multirow{3}{*}{$2$} 
& -- & -- & -- & -0.0790 & -0.0169 \\
& -- & -- & -- & (0.0110) & (0.0061) \\
& -- & -- & -- & [0.0696] & [0.0763] \\

\multirow{3}{*}{$3$} 
& -- & -- & -- & -0.1312 & -0.0381 \\
& -- & -- & -- & (0.0233) & (0.0082) \\
& -- & -- & -- & [0.0783] & [0.0823] \\

\multirow{3}{*}{$5$} 
& -- & -- & -- & -0.1886 & -0.0517 \\
& -- & -- & -- & (0.0434) & (0.0171) \\
& -- & -- & -- & [0.0887] & [0.1206] \\

\multirow{3}{*}{$10$} 
& -- & -- & -- & -0.2722 & -0.0921 \\
& -- & -- & -- & (0.0937) & (0.0483) \\
& -- & -- & -- & [0.1408] & [0.2006] \\

\multirow{3}{*}{$20$} 
& -- & -- & -- & -0.3379 & 0.3865 \\
& -- & -- & -- & (0.1374) & (0.3279) \\
& -- & -- & -- & [0.1530] & [0.4246] \\

\hline \\[-1.8ex]
\end{tabular}
\end{table}

\begin{table}[H]
\caption{Prediction performance on independent test data for the high-dimensional linear setting with 100 repetitions and the spatial error family across different number of locations $K$. Reported are the root mean squared error of prediction (RMSEP), mean absolute error of prediction (MAEP), and quasi negative log-likelihood (NLL) for model-based gradient boosting with gradient boosting (GB-GB) and gradient boosting with deselection (DS-GB).}
\centering
\begin{tabular}{@{\extracolsep{5pt}}clccccc}
\\[-1.8ex] \hline
\hline \\[-1.8ex]
$K$ & Metric & QML & GMM & LS-GB & GB-GB & DS-GB \\
\hline \\[-1.8ex]

\multirow{3}{*}{$1$}
& RMSEP & -- & -- & -- & 1.0670 & 1.0699 \\
& MAEP  & -- & -- & -- & 0.8528 & 0.8550 \\
& NLL   & -- & -- & -- & 852.61 & 861.08 \\

\multirow{3}{*}{$2$}
& RMSEP & -- & -- & -- & 1.0778 & 1.0798 \\
& MAEP  & -- & -- & -- & 0.8606 & 0.8622 \\
& NLL   & -- & -- & -- & 860.74 & 863.89 \\

\multirow{3}{*}{$3$}
& RMSEP & -- & -- & -- & 1.0727 & 1.0740 \\
& MAEP  & -- & -- & -- & 0.8561 & 0.8570 \\
& NLL   & -- & -- & -- & 861.34 & 859.61 \\

\multirow{3}{*}{$5$}
& RMSEP & -- & -- & -- & 1.0718 & 1.0740 \\
& MAEP  & -- & -- & -- & 0.8539 & 0.8554 \\
& NLL   & -- & -- & -- & 862.84 & 860.15 \\

\multirow{3}{*}{$10$}
& RMSEP & -- & -- & -- & 1.0774 & 1.0800 \\
& MAEP  & -- & -- & -- & 0.8567 & 0.8587 \\
& NLL   & -- & -- & -- & 863.85 & 857.20 \\

\multirow{3}{*}{$20$}
& RMSEP & -- & -- & -- & 1.1004 & 1.1088 \\
& MAEP  & -- & -- & -- & 0.8802 & 0.8872 \\
& NLL   & -- & -- & -- & 889.59 & 832.38 \\

\hline \\[-1.8ex]
\end{tabular}
\end{table}

\section{Further case studies} \label{app:case}
\subsection{Boston housing prices}
\begin{figure}[H] 
    \centering
    \includegraphics[width = \textwidth]{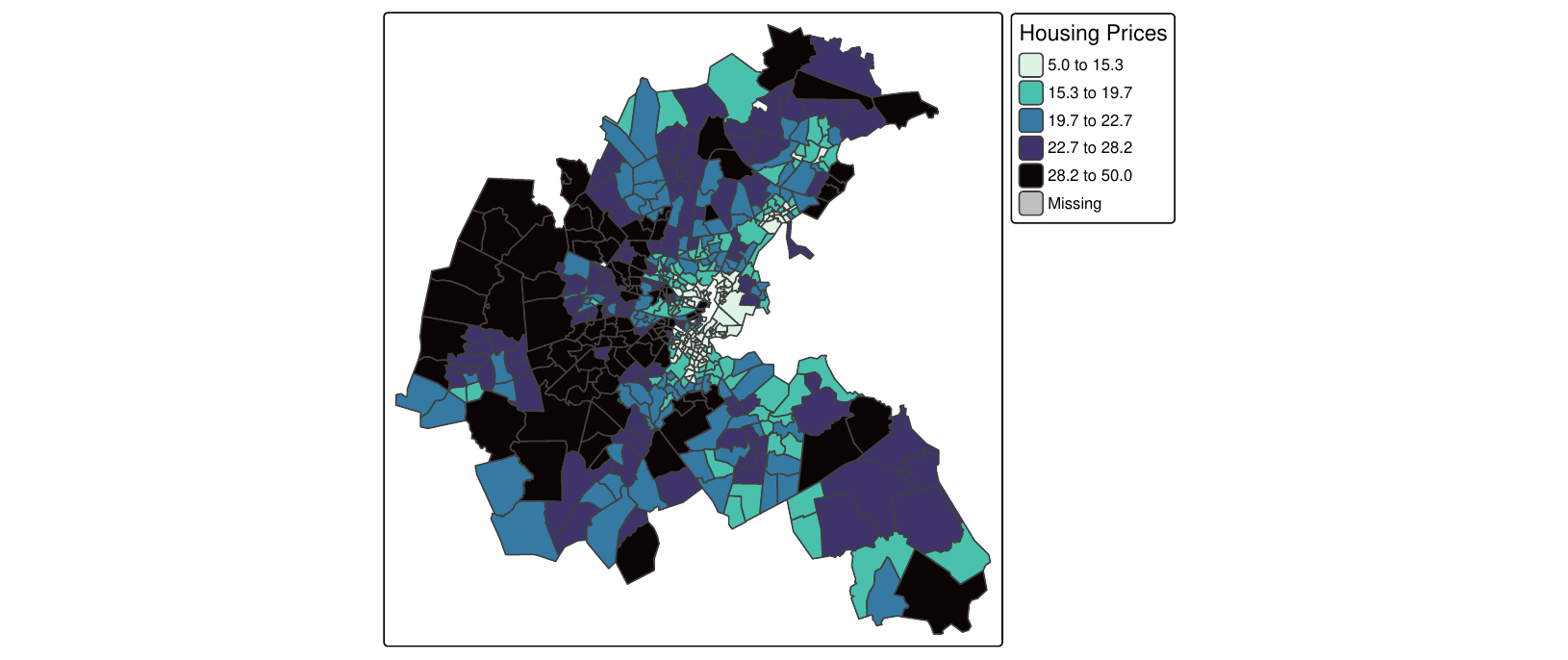}
    \caption{Corrected median value of owner-occupied housing in USD 1000 (CMEDV) in Boston.}
    \label{fig:boston}
\end{figure}

In this section, the classic Boston housing prices data set is revisited \citep{harrison1978}, which is currently available in the \textbf{spData} package for the R programming language \citep{bivand2025}. The dataset comprises 506 observations and 20 independent variables. Descriptions and summary statistics for the relevant variables are presented in Tables \ref{tab:boston} and \ref{tab:boston2}. The objective is to model the corrected median value of owner-occupied housing in thousands of USD (CMEDV). To achieve this, a spatial regression model with autoregressive disturbances, specifically the SDEM, is employed. As illustrated in Figure \ref{fig:boston}, the CMEDV variable exhibits clear spatial clustering justifying the inclusion of spatial dependencies in the modeling process. To maintain consistency with the simulation study in Section \ref{sec:sim}, results are reported for the QML, GMM, LS-GB, GB-GB, DS-GB, and DS-DS estimation strategies. For model-based gradient boosting, the optimal stopping criterion$ m_{\text{opt}}$ is determined via 25-fold subsampling, with a learning rate set to $s = 0.1$. The estimated coefficients for all approaches are shown in Table \ref{tab:bosres}. The results show that the spatial autoregressive parameter $\lambda$ is underestimated in GMM and increases in the model-based gradient boosting algorithm depending on the first step estimation method. Overall, the presence of positive spatial autocorrelation among neighborhoods is confirmed, with $\lambda$ values around $0.6$ across all models. Interestingly, the intercept changes sign, that is, the sign is positive for QML and GMM, but negative when model-based gradient boosting is utilized in the first step. In terms of variable selection, LS-GB eliminates the INDUS variable and most spatial lags of exogenous variables. GB-GB and DS-GB yield nearly identical results, differing only slightly in coefficient estimates. Notably, most spatial lags of exogenous variables are excluded from the final models, effectively simplifying the SDEM toward a SEM in a data-driven manner. In addition to INDUS, the variables NOX and RAD are excluded in the final models, while DS-DS also removes ZN and AGE. This results in a final model with only eight variables which is much more parsimonious compared to the full model that includes $24$ variables. Regarding coefficient interpretation, the algebraic signs remain consistent across all considered estimation strategies. As expected, only the magnitude of the effects varies, suggesting that some variables are more affected by shrinkage in model-based gradient boosting than others. These findings align well with previous research using the same dataset \citep{harrison1978}.

\begin{table}[H]
\caption{Data type and description of dependent and independent variables for Boston housing data.}
\label{tab:boston}
\begin{tabular}{@{\extracolsep{5pt}}llp{9cm}} 
\\[-1.8ex] \hline
\hline \\[-1.8ex]
\textbf{Variable} & \textbf{Type} & \textbf{Description} \\ 
\hline \\[-1.8ex]
CMEDV    & Numeric & Corrected median values of owner-occupied housing in USD 1000 \\
CRIM     & Numeric & Per capita crime rate by town \\
ZN       & Numeric & Proportion of residential land zoned for lots over 25000 sq. ft per town \\
INDUS    & Numeric & Proportion of non-retail business acres per town \\
NOX      & Numeric & Nitric oxides concentration (parts per 10 million) per town \\
RM       & Numeric & Average number of rooms per dwelling \\
AGE      & Numeric & Proportion of owner-occupied units built prior to 1940 \\
DIS      & Numeric & Weighted distances to five Boston employment centres \\
RAD      & Numeric & Index of accessibility to radial highways per town  \\
TAX      & Numeric & Full-value property-tax rate per USD 10,000 per town  \\
PTRATIO  & Numeric & Pupil-teacher ratios per town  \\
B        & Numeric & $1000*(Bk - 0.63)^2$ where $Bk$ is the proportion of blacks by town \\
LSTAT    & Numeric & Percentage of lower status population \\
\hline
\end{tabular}
\end{table}

\begin{table}[H]
\centering
\caption{Summary statistics of dependent and independent variables for Boston housing data.}
\label{tab:boston2}
\begin{tabular}{@{\extracolsep{5pt}}lcccccc} 
\\[-1.8ex] \hline
\hline \\[-1.8ex]
Variable & Mean  & Min & 25\% & Median & 75\% & Max \\
\hline \\[-1.8ex]
CMEDV     & 22.53 & 5.00   & 17.02 & 21.20 & 25.00 & 50.00 \\
CRIM      & 3.61  & 0.00632 & 0.08205 & 0.25651 & 3.67708 & 88.97620 \\
ZN        & 11.36 & 0.00    & 0.00    & 0.00    & 12.50   & 100.00   \\
INDUS     & 11.14 & 0.46    & 5.19    & 9.69    & 18.10   & 27.74    \\
NOX       & 0.555 & 0.3850  & 0.4490  & 0.5380  & 0.6240  & 0.8710   \\
RM        & 6.29  & 3.561   & 5.886   & 6.208   & 6.623   & 8.780    \\
AGE       & 68.57 & 2.90    & 45.02   & 77.50   & 94.08   & 100.00   \\
DIS       & 3.80  & 1.130   & 2.100   & 3.207   & 5.188   & 12.127   \\
TAX       & 408.2 & 187     & 279     & 330     & 666     & 711      \\
PTRATIO   & 18.46 & 12.60   & 17.40   & 19.05   & 20.20   & 22.00    \\
B         & 356.67 & 0.32   & 375.38  & 391.44  & 396.23  & 396.90   \\
LSTAT     & 12.65 & 1.73    & 6.95    & 11.36   & 16.95   & 37.97    \\
\hline \\[-1.8ex]
\end{tabular}
\end{table}

\begin{table}[H]
\centering
\caption{\label{tab:bosres} Coefficient estimates in Boston housing data for quasi-maximum likelihood (QML), generalized method of moments (GMM), model-based gradient boosting with first step OLS
(LS-GB), gradient boosting (GB-GB), gradient boosting with deselection (DS-GB) and gradient boosting with deselection and additional deselection (DS-DS).}
\begin{tabular}{@{\extracolsep{5pt}}lcccccc}
\\[-1.8ex] \hline
\hline \\[-1.8ex]
{} & QML & GMM & LS-GB & GB-GB & DS-GB & DS-DS \\
\hline \\[-1.8ex]
$\lambda$              & 0.6476 & 0.5250 & 0.5250 & 0.5745 & 0.5862 & 0.5862\\
$\text{Intercept}$     & 10.3906 & 23.3001 & 3.8163 & -3.3772 & -3.7616 & -5.7290 \\
$\text{CRIM}$          & -0.0805 & -0.0798 & -0.0537 & -0.0441 & -0.0442 & -0.0526 \\
$\text{ZN}$            & 0.0321 & 0.0344 & 0.0233 & 0.0166 & 0.0166 & \\
$\text{INDUS}$         & -0.0202 & -0.0243 & & & & \\
$\text{NOX}$           & -13.6497 & -14.3729 & -4.2629 & & & \\
$\text{RM}$            & 4.6744 & 4.3191 & 4.6161 & 4.6723 & 4.6834 & 4.7328 \\
$\text{AGE}$           & -0.0437 & -0.0393 & -0.0242 & -0.0206 & -0.0211 & \\
$\text{RAD}$           & 0.2488 & 0.2497 & 0.0317 & & & \\
$\text{DIS}$           & -1.2180 & -1.2170 & -0.9905 & -0.6458 & -0.6277 & -0.5162 \\
$\text{TAX}$           & -0.0140 & -0.0141 & -0.0062 & -0.0053 & -0.0054 & -0.0063 \\
$\text{PTRATIO}$       & -0.4954 & -0.5066 & -0.5300 & -0.4929 & -0.4885 & -0.5320 \\
$\text{B}$             & 0.0127 & 0.0122 & 0.0105 & 0.0100 & 0.0101 & 0.0103 \\
$\text{LSTAT}$         & -0.3158 & -0.3475 & -0.3559 & -0.3457 & -0.3429 & -0.3682 \\
$\bm{W}$\text{CRIM}    & -0.0355 & -0.0343 & & & & \\
$\bm{W}$\text{ZN}      & -0.0047 & 0.0057 & & & & \\
$\bm{W}$\text{INDUS}   & 0.1156 & 0.1152 & & & & \\
$\bm{W}$\text{NOX}     & -1.7040 & -3.3760 & & & & \\
$\bm{W}$\text{RM}      & 2.4427 & 1.3775 & 1.7783 & 1.9820 & 2.0092 & 2.2325 \\
$\bm{W}$\text{AGE}     & 0.0263 & 0.0464 & 0.0047 & & & \\
$\bm{W}$\text{RAD}     & 0.0527 & 0.0825 & 0.0156 & & & \\
$\bm{W}$\text{DIS}     & -0.1823 & -0.2014 & & & & \\
$\bm{W}$\text{TAX}     & -0.0006 & -0.0008 & & & & \\
$\bm{W}$\text{PTRATIO} & -0.3619 & -0.4683 & -0.0510 & & & \\
$\bm{W}$\text{B}       & 0.0003 & -0.0011 & & & & \\
$\bm{W}$\text{LSTAT}   & -0.0282 & -0.1259 & -0.0403 & & & \\
$\sigma$               & 3.5228 & 3.759318 & 3.7360 & 3.7246 & 3.7095 & 3.7459 \\
\hline \\[-1.8ex]
\end{tabular}
\end{table}

\subsection{Columbus crime rate}
\begin{figure}[H] 
    \centering
    \includegraphics[width = \textwidth]{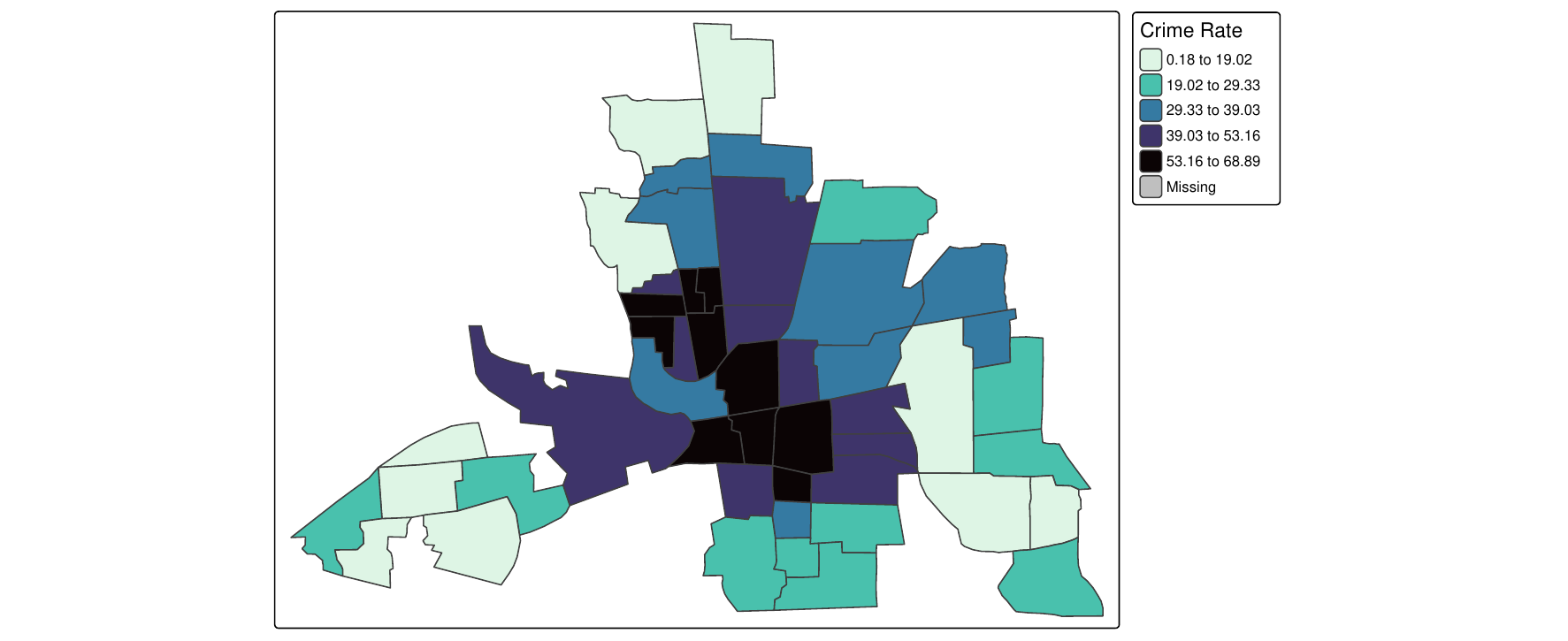}
    \caption{Residential burglaries and vehicle thefts per thousand households in the neighborhood (CRIME) in Columbus.}
    \label{fig:columbus}
\end{figure}

In this section, the classic Columbus crime rate data set is revisited \citep{anselin1988}. The current version of the data is freely available in the \textbf{spData} package in the programming language R \citep{bivand2025}. The Columbus crime rate data is composed of 49 observations and 22 independent variables. Description and summary statistics for the relevant variables are given in \ref{tab:columbus} and \ref{tab:columbus2}. The goal is to model the residential burglaries and vehicle thefts per thousand households in the neighborhood (CRIME). To this end, a spatial regression model with autoregressive disturbances is employed. Particularly, the SDEM is chosen. Indeed, Figure \ref{fig:columbus} reveals clear clustering of the CRIME variable between neighboring locations. Thus, it is realistic to include spatial dependencies when modeling the CRIME variable. To keep a close connection to the simulation study in Section \ref{sec:sim}, results are reported for QML, GMM, LS-GB, GB-GB, DS-GB and DS-DS. For model-based gradient boosting, the optimal stopping criterion $m_{\text{opt}}$ is found via 25-fold subsampling and the learning rate is set to $s = 0.1$. The estimated coefficients for all considered estimation strategies can be seen in Table \ref{tab:colres}. The results for the spatial autoregressive parameter $\lambda$ in QML and GMM indicate almost no or very weak spatial dependence in the error terms. In contrast, model-based gradient boosting constructs parsimonious models, where previously included variables are excluded while increasing $\lambda$. This result is in line with observations in the simulation study. Modeling non-informative variables decreases accuracy in the prediction of the residual vector in the first step, thereby introducing a bias in the estimate of $\lambda$. Since model-based gradient boosting reduces the number of considered non-informative variables, the estimate of $\lambda$ increases. Thus, there seems to be weak spatial autocorrelation between neighboring locations in Columbus. Regarding the variable selection, model-based gradient boosting is able to construct parsimonious final models including only three out of ten variables, namely INC, HOVAL and DISCBD. The algebraic sign and general direction of coefficients remains consistent across all considered estimation strategies although the magnitude changes due to either shrinkage or natural removal of variables. The most important result for the Columbus data is that the data-driven model selection property is clearly illustrated. Particularly, almost all model-based gradient boosting strategies besides LS-GB do not include spatial lags of exogenous variables. Thus, the final model has reduced from the SDEM to a simple SEM model which is the result from starting with an empty model and iteratively adding variables until the optimal stopping criterion is reached. Furthermore, the results are in line with previous research utilizing the same data set \citep{anselin1988}.

\begin{table}[H]
\caption{Data type and description of dependent and independent variables for Columbus neighborhood data.}
\label{tab:columbus}
\begin{tabular}{@{\extracolsep{5pt}}llp{9cm}} 
\\[-1.8ex] \hline
\hline \\[-1.8ex]
\textbf{Variable} & \textbf{Type} & \textbf{Description} \\ 
\hline \\[-1.8ex]
CRIME & numeric &  Residential burglaries and vehicle thefts per thousand households in the neighborhood \\
INC      & Numeric & Household income in thousands of USD \\
HOVAL    & Numeric & Housing value in thousands of USD \\
DISCBD   & Numeric & Distance from the neighborhood to the Central Business District \\
PLUMB    & Numeric & Percentage of housing units without plumbing \\
OPEN     & Numeric & Amount of open space in the neighborhood \\
\hline
\end{tabular}
\end{table}

\begin{table}[H]
\centering
\caption{Summary statistics of dependent and independent variables for Columbus neighborhood data.}
\label{tab:columbus2}
\begin{tabular}{@{\extracolsep{5pt}}lcccccc} 
\\[-1.8ex] \hline
\hline \\[-1.8ex]
Variable & Mean  & Min & 25\% & Median & 75\% & Max \\
\hline \\[-1.8ex]
CRIME    & 35.13 & 0.18   & 20.05  & 34.00  & 48.59  & 68.89 \\
INC      & 14.38 & 4.48   & 9.96   & 13.38  & 18.32  & 31.07 \\
HOVAL    & 38.44 & 17.90  & 25.70  & 33.50  & 43.30  & 96.40 \\
DISCBD   & 2.85  & 0.37   & 1.70   & 2.67   & 3.89   & 5.57 \\
PLUMB    & 2.36  & 0.13   & 0.33   & 1.02   & 2.53   & 18.81 \\
OPEN     & 2.77  & 0.00   & 0.26   & 1.01   & 3.94   & 24.99 \\
\hline \\[-1.8ex]
\end{tabular}
\end{table}

\begin{table}[H]
\centering
\caption{\label{tab:colres} Coefficient estimates in Columbus neighborhood data for quasi-maximum likelihood (QML), generalized method of moments (GMM), model-based gradient boosting with first step OLS (LS-GB), gradient boosting (GB-GB), gradient boosting with deselection (DS-GB) and gradient boosting with deselection and additional deselection (DS-DS).}
\begin{tabular}{@{\extracolsep{5pt}}lcccccc}
\\[-1.8ex] \hline
\hline \\[-1.8ex]
{} & QML & GMM & LS-GB & GB-GB & DS-GB & DS-DS \\
\hline \\[-1.8ex]
$\lambda$              & 0.0659 & 0.0053 & 0.0053 & 0.2035 & 0.1930 & 0.1930 \\
$\text{Intercept}$     & 62.1165 & 61.7707 & 60.2356 & 66.9898 & 72.3338 & 72.3338 \\
$\text{INC}$           & -0.8473 & -0.8549 & -0.7892 & -1.2121 & -1.5173 & -1.5173 \\
$\text{HOVAL}$         & -0.2570 & -0.2557 & -0.1953 & -0.1496 & -0.1468 & -0.1468 \\
$\text{DISCBD}$        & -2.6209 & -2.8443 & -6.8180 & -3.3946 & -3.4197 & -3.4197 \\
$\text{PLUMB}$         & 0.4676  & 0.4632  & 0.1879  & 0.4202  &         &         \\
$\text{OPEN}$          & 0.1192  & 0.1083  &         &         &         &         \\
$\bm{W}$\text{INC}     & -0.1547 & -0.1593 &         &         &         &         \\
$\bm{W}$\text{HOVAL}   & 0.2799  & 0.2933  & 0.3314  &         &         &         \\
$\bm{W}$\text{DISCBD}  & -2.6743 & -2.4168 &         &         &         &         \\
$\bm{W}$\text{PLUMB}   & 0.1791  & 0.2052  &         &         &         &         \\
$\bm{W}$\text{OPEN}    & -0.1557 & -0.2089 &         &         &         &         \\
$\sigma$               & 8.9618  & 8.9716  & 9.1094  & 9.3560  & 9.7158  &  9.7158    \\
\hline \\[-1.8ex]
\end{tabular}
\end{table}

\section{Gradient boosting for spatial error models} \label{app:sem}
Consider following spatial regression model with autoregressive disturbances
\begin{equation*} 
\begin{aligned}
    \bm{y} &= \bm{X}\bm{\beta}  + \bm{u} \\
    \bm{u} &= \lambda\bm{W}\bm{u} + \bm{\epsilon}
\end{aligned}
\end{equation*}
which can be identified as a simple SEM. In principle, spatial dependence enters this model only through the disturbances and not via a spatial lag of exogenous variables as in the SDEM. The SEM is a special case of the SDEM, which can be easily recovered by removing the spatial lags of exogenous variables from the SDEM. Thus, the SEM is nested under the SDEM. This implies that the loss function and the negative gradient vector remain identical as in the SDEM which are given by Equations \ref{eq:loss} and \ref{eq:grad}. In principle, model-based gradient boosting can then be performed via Algorithm \ref{algo:boost} and deselection via Algorithm \ref{algo:des}. However, the feasibility still relies on the knowledge about the spatial autoregressive parameter $\lambda$. Thus, a feasible model-based gradient boosting algorithm can be obtained by replacing the loss function and the negative gradient vector by the expressions given in Equations \ref{eq:ende} and \ref{eq:ende2}.

\section{Gradient boosting for cross-regressive models} \label{app:slx}
Consider following spatial regression model
\begin{equation*}
    \bm{y} = \bm{X}\bm{\beta} + \bm{W} \bm{X} \bm{\theta} + \bm{\epsilon}
\end{equation*}
which can be identified as the SLX. In principle, spatial dependence enters this model only through the spatial lag of exogenous variables as in the SDEM. The SLX can be understand as a special case of the SDEM by removing the autoregessive nature of the disturbances. However, the principle model structure is severely different which is also evident that the loss function and the negative gradient vector are not identical to the SDEM and SEM. However, the structure of the model is far simpler than in SDEM or SEM such that the coefficients can be easily estimated by OLS. In turn, model-based gradient boosting does not rely on knowledge about the spatial autoregressive parameter $\lambda$ and thus can easily be applied by adjusting the corresponding loss function. Particularly, the loss function becomes the simple squared error loss given by 
\begin{equation*}
    \rho(\bm{y}, \bm{\eta})  = (\bm{y} - \bm{\eta})^{\prime} (\bm{y} - \bm{\eta}).
\end{equation*}
The negative gradient vector is then obtained as 
\begin{equation*}
    -\frac{\partial}{\partial \bm{\eta}}\rho(\bm{y}, \bm{\eta})  =  2(\bm{y} - \bm{\eta})
\end{equation*}
where $\bm{\eta} = (\bm{X}, \bm{W} \bm{X}) (\bm{\beta}, \bm{\theta})^{\prime} = \bm{Z} \bm{\delta}$. Model-based gradient boosting can thus be directly performed by Algorithm \ref{algo:boost} and deselection by Algorithm \ref{algo:des} if corresponding loss function and negative gradient vector are replaced. Additionally, model-based gradient boosting with the squared error loss yields consistent estimators of the coefficients of interest. Results for convergence and consistency properties in low- as well as high-dimensional linear settings are given in \cite{zhang2005} and \cite{bühlmann2006}.

\end{appendices}

\bibliography{sn-bibliography}

\end{document}